\documentclass[aps,prb,twocolumn,showpacs,superscriptaddress,floatfix,citeautoscript]{revtex4-1}
\usepackage{times}
\usepackage{graphicx}
\usepackage{dcolumn}
\usepackage{bm}
\usepackage{color}
\begin{document}

\title{Fate of disorder-induced inhomogeneities in strongly correlated d-wave superconductors}

\author{Debmalya Chakraborty}
\affiliation{Indian Institute of Science Education and Research-Kolkata, Mohanpur Campus, India-741252}
\author{Amit Ghosal}
\affiliation{Indian Institute of Science Education and Research-Kolkata, Mohanpur Campus, India-741252}

\begin{abstract}
We analyze the complex interplay of the strong correlations and impurities in a high temperature superconductor and show that both the nature and degree of the inhomogeneities at zero temperature in the local order parameters change drastically from what are obtained in a simple Hartree-Fock-Bogoliubov theory.
While both the strong electronic repulsions and disorder contribute to the nanoscale inhomogeneity in the population of charge-carriers, we find them to compete with each other leading to a relatively smooth variation of the local density.
Our self-consistent calculations modify the spatial fluctuations in the pairing amplitude by suppressing all the double-occupancy within a Gutzwiller formalism and prohibit the formation of distinct superconducting-`islands'.
In contrast, presence of such `islands' controls the outcome if strong correlations are neglected.
The reorganization of the spatial structures in the Gutzwiller method makes these superconductors surprisingly insensitive to the impurities.
This is illustrated by a very weak decay of superfluid stiffness, off-diagonal long range order and local density of states up to a large disorder strength.
Exploring the origin of such a robustness we conclude that the underlying one-particle normal states reshape in a rich manner, such that the superconductor formed by pairing these states experiences a weaker but spatially correlated effective disorder.
Such a route to superconductivity is evocative of Anderson's theorem.
Our results capture the key experimental trends in the cuprates.
\end{abstract}

\maketitle

\section{Introduction}\label{sec:Intro}

The study of disordered high-temperature superconductors (HTSC) \cite{Hirschfeld09,Balatsky06} is scientifically rewarding for multiple reasons. First, we learn to deal with strongly correlated electrons. Significance of strong repulsive correlations is paramount in these superconductors which make their parent (undoped) compound an antiferromagnetic Mott insulator \cite{Sudip89}. Second, the superconductivity in the HTSC cuprates is believed to originate from the two-dimensional copper-oxide (${\rm CuO_2}$) planes \cite{Millis00}, and is rife with complex effects of enhanced quantum fluctuations from the reduced dimensionality \cite{Mermin66,Hohenberg67}. Finally, these systems provide a hotbed for the complex interplay of electronic interactions and disorder \cite{Dagotto05}, both being strong.
By tuning various parameters describing these systems including disorder, the cuprates can be made to undergo quantum phase transitions \cite{SubirBook00} to various non-superconducting phases. Each of such quantum phase transitions can vastly add to our knowledge of the condensed phases of matter.

While the physics of HTSC is far from being settled for the unusual normal state, the superconductivity in the clean systems is well described by a BCS-like ground state (GS) \cite{Hussey02} at low temperatures ($T$). Such a BCS-GS is identified with d-wave pairing amplitude \cite{Shen93,Ding95,Shen03}, and hence they are referred to as d-wave superconductors (dSC). The corresponding Bogoliubov quasiparticles result into a linear low-lying density of states (DOS) \cite{Maki94}. The effect of impurities on this BCS-GS was developed within the conventional Abrikosov-Gorkov theory \cite{Abrikosov61}, which infers that the non-magnetic impurities easily destroy the d-wave superconductivity \cite{Maki01}. Contrary to this wisdom, the cuprates are found significantly robust to disorder in the following senses: (1) The local density of states (LDOS) maintains the same low-lying V-shaped gap as in a pure dSC even in the presence of the disorder \cite{Cren01,Kapitulnik06,Yazdani08}.
(2) Even though the gap-map develops nanoscale inhomogeneities \cite{Pan01,Lang02}, the low-energy LDOS remains surprisingly homogeneous \cite{McElroy05}. (3) The superfluid density undergoes only a modest reduction with the impurities \cite{Nachumi96}.
(4) The superconducting gap in the nodal region gets suppressed by disorder while an enhancement of the antinodal gap is found \cite{Hashimoto09} -- a signature contributing to the nodal-antinodal dichotomy \cite{Zhou04,Shen05}. (5) Scanning tunnelling spectroscopy (STS) yielded an intriguing energy-resolved nanoscale density modulation persisting on top of the inhomogeneous background at low $T$, as signalled by Fourier transformed local density of states (FT-LDOS) \cite{McElroy05,Howald01,Jenny02}. While much of these surprises have been attributed to the strong electronic repulsions in these materials, a systematic calculation including both the correlations and disorder \cite{Tanskovic03}, had always been challenging. The first step towards integrating strong correlations and disorder in a numerical calculation \cite{Garg08} has already uncovered interesting effects.

We incorporate both the strong interactions and disorder on the same footing for a two-dimensional dSC, and our main results are summarized as follows:
(1) The nature and degree of the inhomogeneities in the order parameters change substantially from the scenario that neglects strong correlations. In particular, superconducting-`islands' cannot be discerned even at large disorder, though the nanoscale inhomogeneities develop.
(2) Superfluid density, off-diagonal long range order and density of states show amazing insensitivity to the impurities. (3) The interplay of the strong correlations and bare disorder produces an effective background that is {\it correlated} in space.
(4) Our results offer a simple description of the complex problem of disordered cuprates motivating a pairing mechanism similar to Anderson's theorem \cite{Anderson59}, which traditionally describes only the s-wave superconductors (sSC).

Significant progress had been made in the past addressing separately the effects of disorder \cite{Balatsky06}, or strong correlations \cite{Zhang88} on a dSC. The robustness of dSC to impurities had been partially addressed \cite{Ghosal00,Hirschfeld00} within the Bogoliubov de-Gennes (BdG) mean-field theory that includes the inhomogeneities in the {\it local} order parameters, however, ignores any phase-space restrictions \cite{Vollhardt84} from the strong correlations. We broadly refer to such calculations as inhomogeneous mean-field theory (IMT).
The IMT produces an inhomogeneous pairing \cite{Pan01,Lang02}, a gradual fall of the superfluid density with disorder \cite{Ghosal00,Franz97}, a weaker filling of the low-lying DOS \cite{Ghosal00,Hirschfeld00} and it helps visualizing the proposal for a `swiss-cheese' \cite{Nachumi96,Stroud01} model. The IMT calculations had been extended to study the responses of a dSC to antiferromagnetism \cite{Andersen07}, competing orders \cite{Ghosal05,Schmid13} and vortex-lattice \cite{Ghosal02,Ting01}.
However, the significance of the IMT results for HTSC is not obvious because of its mean-field treatment of the effective interactions, leaving the connection between the {\it bare} strong correlations and superconducting pairing artificial, and possibly dubious. 
The prime role of strong repulsive correlations in clean (pure) cuprates lies in prohibiting the double-occupancy of the charge carriers. A comprehensive theory for such a constrained GS is based on Gutzwiller's projected wave functions \cite{Gutzwiller63,Anderson87,Arun04,Anderson04}. However, a simple framework, called Gutzwiller approximation (GA) \cite{Fazekasbook99}, that restricts all double-occupancy by renormalizing the bare parameters of the Hamiltonian, had also been developed. The renormalization is easily rationalized, e.g., the electronic hopping is difficult due to the prohibition of double-occupancy, while the effects of exchange interactions are enhanced due to a large number of singly occupied sites.
The resulting model is solved within the framework of a renormalized mean-field theory \cite{Zhang88,Wang06,Chen08,Huang05},
which describes clean strongly correlated cuprates reasonably well \cite{Kotliar88,Rajdeep07}.

Several attempts have been made recently to extend the GA analysis to the inhomogeneous superconductors \cite{Edegger07, Ko07, Fukushima08, Himeda99,Ogata02}. We use the notation GIMT, for referring to such IMT calculations augmented with inhomogeneous GA.
Incorporating the strong correlations, it has been shown \cite{Garg08, Hirschfeld11, Fukushima09} that the robustness of superconductivity goes far beyond what is obtained from a simple IMT. In particular, the site-averaged low-energy DOS remains nearly unaffected up to a rather large concentration ($\sim 25\%$) of impurities \cite{Garg08}! Such striking observation raises several questions: (1) How does the strong correlation make the impurities essentially disappear from the low-energy physics? (2) What happens to the disorder-induced inhomogeneities, whose presence is instrumental \cite{Ghosal00, Ghosal01} for the inferences made by the simple IMT calculations? And finally, (3) How do the other observables, e.g., superfluid density or off-diagonal long range order (ODLRO) respond to the disorder?
Answers to these questions are necessary for a better understanding of dirty superconductivity and we address them in this paper. Importance of inhomogeneous GA are currently being probed in a variety of contexts, e.g., on competing charge and spin stripes \cite{Sigrist09}, spin-density wave \cite{Hirschfeld11}, vortex-core physics \cite{Ogata03}, antiphase superconducting domain, and on the moment formation around a single impurity \cite{Tsuchiura01}.

The rest of the paper is organized as follows.
In section~\ref{sec:Frwork}, we describe the model and parameters used for our study and also review the GA formalism for inhomogeneous systems. We discuss our results, together with a quantitative comparison between the IMT and GIMT schemes in section~\ref{sec:results}, keeping our focus on the behavior of the physical observables.
We further study the spatial structures in all the order parameters. We establish a connection between the nature of the inhomogeneities and the behavior of observables. In addition, we illustrate the importance of the underlying normal states in comprehending our findings. Finally, we summarize and conclude in section~\ref{sec:conclu}.

\section{Theoretical framework}\label{sec:Frwork}

\subsection{Strong correlations and Gutzwiller approximation for an inhomogeneous d-wave superconductor}\label{subsec:GA}

We describe the phases of the strongly correlated cuprates by the Hubbard model, which is a minimal lattice model representing repulsively interacting band electrons,
\begin{equation}
{\cal H}_{\rm Hubb}=\sum_{ ij \sigma} t_{ij} (c^{\dagger}_{i \sigma} c_{j \sigma}+h.c.)+U\sum_j n_{j \uparrow}n_{j \downarrow}.
\label{eq:Hubb}
\end{equation}
Here, the first term indicates electrons' hopping on a two-dimensional square lattice, with $c^{\dagger}_{i \sigma}(c_{i \sigma})$ as the creation (annihilation) operator of the electron on the site $i$ and with spin $\sigma$. We take $t_{ij}=-t$, when $i$ and $j$ are nearest neighbors, denoted as $\langle ij \rangle$, and  $t_{ij}=t'$, when $i$ and $j$ are next-nearest neighbors, with the notation of $\langle \langle ij\rangle \rangle$. We choose $t_{ij}=0$ for all other pairs of $i$ and $j$. $U$ is the onsite interaction energy between electrons occupying the same site and mimics a Coulomb-type repulsion where inter-site contributions are neglected in favor of screening. The local density operator with spin $\sigma$ at site $j$ is denoted as $n_{j \sigma}=c^{\dagger}_{j \sigma} c_{j \sigma}$.

In the strongly correlated limit $U\gg t$, the Schrieffer-Wolff transformation on ${\cal H}_{\rm Hubb}$ yields an effective $``t-J"$ model \cite{Spalek77}:
\begin{eqnarray}
{\cal H}_{\rm ``t-J"} &=& \sum_{ij \sigma}t_{ij} (\tilde{c}^{\dagger}_{i \sigma} \tilde{c}_{j \sigma}+h.c.)+\sum_{ij}J_{ij}\Big(\tilde{\mathbf{S}}_i.\tilde{\mathbf{S}}_j-\frac{\tilde{n}_i\tilde{n}_j}{4}\Big) \nonumber \\
&-&\frac{J}{4}\sum_{\langle ijm \rangle, \sigma \atop m\neq i} (\tilde{c}^{\dagger}_{i \sigma} \tilde{n}_{j \bar{\sigma}} \tilde{c}_{m \sigma}-\tilde{c}^{\dagger}_{i \sigma} \tilde{c}^{\dagger}_{j \bar{\sigma}} \tilde{c}_{j \sigma} \tilde{c}_{m \bar{\sigma}}+h.c.) \nonumber \\
&-&\frac{J'}{4}\sum_{\langle\langle ijm \rangle\rangle, \sigma \atop m\neq i} (\tilde{c}^{\dagger}_{i \sigma} \tilde{n}_{j \bar{\sigma}} \tilde{c}_{m \sigma}-\tilde{c}^{\dagger}_{i \sigma} \tilde{c}^{\dagger}_{j \bar{\sigma}} \tilde{c}_{j \sigma} \tilde{c}_{m \bar{\sigma}}+h.c.) \nonumber \\
&+&\frac{2tt'}{U}\sum_{\sigma, \langle ij \rangle \atop \langle\langle jm \rangle\rangle } (\tilde{c}^{\dagger}_{i \sigma} \tilde{n}_{j \bar{\sigma}} \tilde{c}_{m \sigma}-\tilde{c}^{\dagger}_{i \sigma} \tilde{c}^{\dagger}_{j \bar{\sigma}} \tilde{c}_{j \sigma} \tilde{c}_{m \bar{\sigma}}+h.c.), \nonumber \\
\label{eq:tJ}
\end{eqnarray}
where all terms up to ${\cal O}(t^2/U)$ are kept, and $J_{ij}=4t_{ij}^2/U$. Here, we used the notation $J_{ij}=J$ for $\langle ij \rangle$, and $J_{ij}=J'$ for $\langle \langle ij \rangle \rangle$, and $\tilde{c}_{i\sigma}=c_{i \sigma}(1-n_{i\bar{\sigma}})$ is the annihilation operator in the `projected space' prohibiting double-occupancy at the site $i$. The terms appearing in the second and third line of equation~(\ref{eq:tJ}) represent the `three-site terms', which along with $J'$-term are often neglected in the standard $t-J$ model, even though they are of the same order as the $J$-term. In fact, these terms are capable of generating additional broken symmetry phases in the Cooper-channel, as we discuss in section~\ref{subsec:BdG}. We used the quotes in $``t-J"$ to emphasize that it has contributions beyond the standard $t-J$ model.
The notations $\langle ijm \rangle$ and $\langle\langle ijm \rangle\rangle$ are for the $3$ sites with ($i$, $j$) and ($j$, $m$) being nearest neighbors and next-nearest neighbors to each other respectively.

We introduce disorder by redefining ${\cal H}_{\rm ``t-J"}$ to ${\cal H}_{\rm ``t-J"}+ \sum_{i\sigma} (V_i-\mu)n_{i\sigma}$, where $V_i$ is the (non-magnetic) impurity potential at site $i$ and $\mu$ is the chemical potential that fixes the average density of electrons in the system to a desired value. There are different ways of choosing $V_i$'s, depending on the specifics of the experimental realizations.
It is, however, important to quantify the {\it degree} of disorder by a small number of parameters defining the distribution of $V_i$, for an appropriate comparison with the experiments.

For our calculations, we used three models of disorder:
(i) {\it Box-disorder} : $V_i$'s are drawn from a `box' distribution, such that,  $V_i \in [-V/2,V/2]$ uniformly,
(ii) {\it Concentration disorder} (or {\it conc-disorder}) : $V_i=V_0$ on a given $n_{\rm imp}$ fraction of the sites, which are randomly chosen, and $V_i=0$ for all other sites. 
(iii) {\it Out-of-plane disorder}, as detailed in the supplementary material. 
 While the box-disorder is quantified by a single parameter $V$, the conc-disorder needs two parameters for its characterization: $V_{0}$ and $n_{\rm imp}$. 
We will focus, in particular, on the box-disorder which is better suited to address several conceptual issues as we will see in section~\ref{subsec:Spatial}. All data are averaged over 10-15 independent realizations for a given strength of disorder.

Strong electronic repulsions restrict the Hilbert space of ${\cal H}_{\rm ``t-J"}$ by eliminating all double-occupancies. One simple method to implement such restriction is the GA, in which the higher order off-site correlations are neglected \cite{Fazekasbook99}. Following Ref. \onlinecite{Fukushima08}, we consider the ground-state wave function in the projected space $|\psi\rangle = P|\psi_0\rangle$, where $|\psi_0\rangle$ is the ground-state wave function in the Hilbert space that allows
double-occupancy and $P$ is the projection operator, $P=\prod_iP_i$, with $P_i=\gamma^{{n_i}/2}_i(1-n_{i \uparrow} n_{i \downarrow})$. Here, $\gamma_i$ are the local fugacity factors obtained by demanding conservation of the local electron densities, $\rho_{i\sigma}=\langle n_{i\sigma} \rangle=\langle n_{i\sigma} \rangle_0$.
In GA, the effects of projection on impure dSC are absorbed by local Gutzwiller renormalization factors \cite{Ko07}. The expectation value of a general operator $\hat{A}$ in different ground-state wave functions can be written as ${\langle \hat{A} \rangle}_0 =  {\langle\psi_0|\hat{A}|\psi_0\rangle}/{\langle\psi_0|\psi_0\rangle}$ and ${\langle \hat{A} \rangle} =  {\langle\psi|\hat{A}|\psi\rangle}/{\langle\psi|\psi\rangle}$, and GA amounts to the following simple relations:
\begin{equation}
\langle c^{\dagger}_{i \sigma} c_{j \sigma} \rangle \approx g^t_{ij}\langle c^{\dagger}_{i \sigma} c_{j \sigma} \rangle_0,
\label{eq:gut1}
\end{equation}
\begin{eqnarray}
\langle c^{\dagger}_{i \sigma} n_{j \bar{\sigma}} c_{m \sigma} \rangle &\approx& g^{31}_{ijm}\langle c^{\dagger}_{i \sigma} n_{j \bar{\sigma}} c_{m \sigma} \rangle_0, \nonumber \\
\langle c^{\dagger}_{i \sigma} c^{\dagger}_{j \bar{\sigma}} c_{j \sigma} c_{m \bar{\sigma}} \rangle &\approx& g^{32}_{ijm} \langle c^{\dagger}_{i \sigma} c^{\dagger}_{j \bar{\sigma}} c_{j \sigma} c_{m \bar{\sigma}} \rangle_0, 
\label{eq:gut3}
\end{eqnarray}
\begin{equation}
\langle \mathbf{S}_i.\mathbf{S}_j \rangle \approx g^{xy}_{ij}\langle \mathbf{S}_i.\mathbf{S}_j \rangle_0, {~\rm however,~} \langle n_i n_j \rangle \approx \langle n_i n_j \rangle_0.
\label{eq:gut2}
\end{equation}
Here $g^t_{ij}$, $g^{31}_{ijm}$, $g^{32}_{ijm}$ and $g^{xy}_{ij}$ are the Gutzwiller renormalization factors (GRF) which depend on an appropriate combinations of local doping $x_i, x_j, x_m$ (here $x_i=1-\rho_i$). The GRFs for our system (without any magnetic order) are given by,
\begin{eqnarray}
g^t_{ij}=\sqrt{\frac{4x_ix_j}{(1+x_i)(1+x_j)}}&,& ~~g^{xy}_{ij}=\frac{4}{(1+x_i)(1+x_j)}, \nonumber \\
g^{31}_{ijm}=g^t_{im}&,& ~~g^{32}_{ijm}=\frac{2g^t_{im}}{(1+x_{j})}
\label{eq:gut6}
\end{eqnarray}
We emphasize that since the GRFs are {\it different} for $\mathbf{S}_i.\mathbf{S}_j$ (renormalized by $g^{xy}_{ij}$) and $n_in_j$ (not renormalized) terms in ${\cal H}_{\rm ``t-J"}$, a flat renormalization of the exchange coupling, $J$, would not suffice.

Upon complete removal of the double-occupancy by GA, the total energy of the system can be calculated in terms of the relevant local order parameters: $\rho_i=\sum_{\sigma} \langle n_{i \sigma} \rangle \equiv \sum_{\sigma} \langle n_{i \sigma} \rangle_0$, $\Delta_{ij} \equiv \langle c_{j \downarrow} c_{i \uparrow} \rangle_0 + \langle c_{i \downarrow} c_{j \uparrow} \rangle_0$, and $\tau_{ij} \equiv \langle c_{i \downarrow}^{\dagger} c_{j \downarrow} \rangle_0 \equiv \langle c_{i \uparrow}^{\dagger} c_{j \uparrow} \rangle_0$. We minimize $\langle\psi_0|{\cal H}_{\rm ``t-J"}|\psi_0\rangle$, with respect to $|\psi_0\rangle$ with additional constraint $\langle\psi_0|\psi_0\rangle = 1$ and $\sum_i \rho_i=\rho$ ($\rho$ being the chosen total density), leading to the minimization of the functional $W$, where,
\begin{equation}
W=\langle\psi_0|{\cal H}_{\rm ``t-J"}|\psi_0\rangle -\beta (\langle\psi_0|\psi_0\rangle-1) - \mu (\sum_i \rho_i-\rho ),
\label{eq:energy}
\end{equation}
employing the variational relation $(\delta W)/(\delta \langle \psi_0|)=0$. Here $\beta$ and $\mu$ are the Lagrangian multipliers for the constraints mentioned above. Upon minimizing $W$, we obtain the following BdG mean-field Hamiltonian \cite{Wang06,Hirschfeld11,Fukushima09,Sigrist09},
\begin{eqnarray}
H_{\rm MF}&=&\sum_{ij \sigma} \frac{\partial W}{\partial \tau_{ij}} c^{\dagger}_{i \sigma} c_{j \sigma} + \sum _{ij} \frac{\partial W}{\partial \Delta_{ij}} (c^{\dagger}_{i \uparrow} c^{\dagger}_{j \downarrow}+c^{\dagger}_{j \uparrow} c^{\dagger}_{i \downarrow}) \nonumber \\ 
&+& \sum_{i \sigma} \frac{\partial W}{\partial \rho_{i \sigma}} n_{i \sigma},
\label{eq:meanfield1}
\end{eqnarray}
satisfying the eigenvalue equation: $H_{\rm MF}|\psi_0\rangle=\beta |\psi_0\rangle$. Upon expanding the derivatives in equation~(\ref{eq:meanfield1}), we obtain the explicit form of the mean-field Hamiltonian,
\begin{eqnarray}
{\cal H}_{\rm MF}  &=& \sum_{i, \delta, \sigma}{}^{'} \{ t_{i\delta} g^t_{i,i+\delta}-W^{\rm FS}_{i\delta} \} ~c^{\dagger}_{i \sigma} c_{i+\delta \sigma} \nonumber \\
&+&\sum_{i, \delta}{}^{'}\{G_1^{i,\delta} \Delta_i^{\delta}(c^{\dagger}_{i\uparrow} c^{\dagger}_{i+\delta \downarrow}+c^{\dagger}_{i+\delta \uparrow} c^{\dagger}_{i \downarrow})+h.c.\} \nonumber \\
&+&\sum_{i, \delta, \delta'}{}^{'}\{G_2^{i,\delta,\delta'} \Delta_i^{\delta}(c^{\dagger}_{i\uparrow} c^{\dagger}_{i+\delta' \downarrow}+c^{\dagger}_{i+\delta' \uparrow} c^{\dagger}_{i \downarrow})+h.c.\} \nonumber \\
&+&\sum_{i, \delta, \tilde{\delta}}\{G_3^{i,\delta,\tilde{\delta}} \Delta_i^{\tilde{\delta}}(c^{\dagger}_{i\uparrow} c^{\dagger}_{i+\delta \downarrow}+c^{\dagger}_{i+\delta \uparrow} c^{\dagger}_{i \downarrow})+h.c.\}\nonumber \\
&+&\sum_{i, \delta, \tilde{\delta}}\{G_3^{i,\delta,\tilde{\delta}} \Delta_i^{\delta}(c^{\dagger}_{i\uparrow} c^{\dagger}_{i+\tilde{\delta} \downarrow}+c^{\dagger}_{i+\tilde{\delta} \uparrow} c^{\dagger}_{i \downarrow})+h.c.\} \nonumber \\
&+&\sum_{i, \delta \atop \delta', \sigma}{}^{'}G_4^{i,\delta,\delta'}c^{\dagger}_{i+\delta \sigma} c_{i+\delta' \sigma} + \sum_{i,\sigma} (V_i-\mu+\mu_i^{\rm HS}) n_{i\sigma} \nonumber \\
&+& \sum_{i, \delta, \tilde{\delta}, \sigma}G_5^{i,\delta,\tilde{\delta}} \{ c^{\dagger}_{i+\delta \sigma} c_{i+\tilde{\delta} \sigma} + h.c. \} 
\label{eq:meanfield2}
\end{eqnarray}
We denote the Cooper-channel order parameters $\Delta_{ij}$ on the bond connecting sites $i$ and $j$ as $\Delta_{i}^{\delta}$ where $j=i+ \delta$, with $\delta=\pm x ~{\rm or} \pm y$, $\tilde{\delta}=\pm (x \pm y)$, and the primed-summation means:
\begin{equation}
\sum_{i,\delta,\sigma}{}^{'}\equiv \sum_{i,\delta,\sigma}+\sum_{i,\tilde{\delta},\sigma}~{\rm and}~\sum_{i,\delta,\delta'}{}^{'}\equiv \sum_{i,\delta,\delta'}+\sum_{i,\tilde{\delta},\tilde{\delta}'} \nonumber 
\label{eq:sum}
\end{equation}
The coefficients $G_1$ to $G_5$ in equation~(\ref{eq:meanfield2}) are built with the GRFs defined in equation~(\ref{eq:gut6}) and their exact expressions are included in the \ref{sec:appendix}. $\mu_i^{\rm HS}$ and $W^{\rm FS}_{i\delta}$ are the Hartree- and Fock-shift terms arising from the mean-field decomposition of interactions and are also given in the \ref{sec:appendix}. We found that the major contributions of $\mu_i^{\rm HS}$ come from the derivatives of the GRF for the parameters describing cuprates.

\subsection{Iterative self-consistency: BdG method}\label{subsec:BdG}

We diagonalize ${\cal H}_{\rm MF}$ using the BdG transformations \cite{Gennesbook99}, $c_{i \sigma}=\sum_{n} ( \gamma_{n, \sigma} u_{i,n}-\sigma \gamma^{\dagger}_{n, -\sigma} v_{i,n}^{*}  )$, where $\gamma^{\dagger}_{n,\sigma}$ and $\gamma_{n,\sigma}$ are the creation and annihilation operators of the Bogoliubov quasiparticles. The resulting eigen-system is then solved self-consistently for all the local order parameters. Inclusion of $J'$ and the three-site terms in the Hamiltonian, allows additional Cooper-channel orders in ${\cal H}_{\rm MF}$ , namely; $s_{xy}$ and  $d_{xy}$, in addition to the standard d-wave ($d_{x^2-y^2}$) and the extended s-wave ($s_{xs}$) which take the following forms in the clean systems:
\begin{eqnarray}
\Delta_{d_{x^2-y^2}}({\vec k})&=&\Delta^{(0)}_{d_{x^2-y^2}} ({\rm cos}k_x - {\rm cos}k_y)/2, \\
\Delta_{s_{xs}}({\vec k})&=&\Delta^{(0)}_{s_{xs}} ({\rm cos}k_x + {\rm cos}k_y)/2, \\
\Delta_{s_{xy}}({\vec k})&=&\Delta^{(0)}_{s_{xy}} ({\rm cos}k_x{\rm cos}k_y), \\
\Delta_{d_{xy}}({\vec k})&=&\Delta^{(0)}_{d_{xy}} ({\rm sin}k_x{\rm sin}k_y),
\end{eqnarray}
where, $\Delta^{(0)}_{d_{x^2-y^2}}$, $\Delta^{(0)}_{s_{xs}}$, $\Delta^{(0)}_{s_{xy}}$ and $\Delta^{(0)}_{d_{xy}}$ are the uniform values of the pairing amplitudes with respective symmetries. We denote $\Delta_{d_{x^2-y^2}}$ by $\Delta_d$ for the rest of the paper for notational simplification.
Starting with initial guesses for all the local order parameters, we numerically find out the eigenvalues and eigenfunctions of the BdG Hamiltonian. With these, we recalculate the order parameters and compare them with the initial guesses. If the two do not match on all the sites, the whole process is iterated with a new choice of the order parameters, until self-consistency is achieved. 

\subsection{Model and parameters}\label{subsec:modpars}

We investigated the physics of ${\cal H}_{\rm MF}$ within the framework of GIMT for a wide range of parameters. We will describe our results for $T=0$ in the next section for $U=12$ and $t'=1/4$ in the units of $t$ for all our GIMT calculations, which are believed to describe a typical cuprate \cite{Campuzano95}. 
All other energies are also expressed in the units of $t$. 
Typical size of the unit cell for our calculations had been $30\times 30$, and we focus on the results for the average density of electrons, $\rho=0.8$ which coincides with near optimal doping in the cuprates. In general, HTSC materials show a low temperature symmetry breaking primarily in the Cooper-channel at the optimal doping and hence we do not consider other orders \cite{Sudip01,Fradkin10,Emery99,Subir09,Varma06}. For the box-disorder we tune the disorder strength up to $V=3$, which is sufficient to destroy dSC in the IMT calculations, as we will see in section~\ref{subsec:PhysObs}. The chosen strength of the conc-disorder is $V_0=1$ for most of our calculations setting them to the `Born scattering limit'. While we report some results even for $V_0\sim 5-10$ to address the effect of strong scatterers, we avoid the unitary limit \cite{Hirschfeld11}, e.g., in the case of substitutionary Zinc impurities \cite{Pan99}. They arguably produce a local  antiferromagnetic surrounding \cite{Hirschfeld11}, and can lead to effects beyond the scope of our study. In order to carry out an appropriate comparison, a value of $U=3.69$ is chosen for the IMT calculations which yields the same homogeneous ($V=0$) d-wave gap from the GIMT scheme with the cuprate parameters. An alternative IMT scheme could also be set up and is discussed in the supplementary material. The ${\cal H}_{\rm MF}$ for plain IMT is recovered by setting the GRF in equation~(\ref{eq:gut6}) to unity, however, suffers from a technical problem. The three-site terms deplete the $d_{x^2-y^2}$ order heavily leading to a cancellation of much of the d-wave pairing amplitudes. The GIMT calculations, however, remains free from such a deficit. Further, the $s_{xy}$ and $d_{xy}$ orders from the $J'$ term were found insignificant in the GIMT calculations, as will be illustrated in figure~\ref{fig:Pden}(c). Thus, we present bulk of our results using just the $t-t'-J$ model for the GIMT calculations with box-disorder and for IMT calculations with both the box- and the conc-disorder, dropping the three-site terms and $J'$ term in favor of simplicity. Our GIMT results with conc-disorder allow all the Cooper-channel orders as it uses the full $``t-J"$ model.

\section{Results}\label{sec:results}

We begin this section showing our results quantifying the nature and degree of the inhomogeneities in the local order parameters in both the GIMT and plain IMT calculations.

\subsection{Distribution of the order parameters}\label{subsec:Pdel}

\begin{figure*}
\begin{center}
\begin{minipage}{.9\columnwidth}
\includegraphics[width=1.0\columnwidth,clip=]{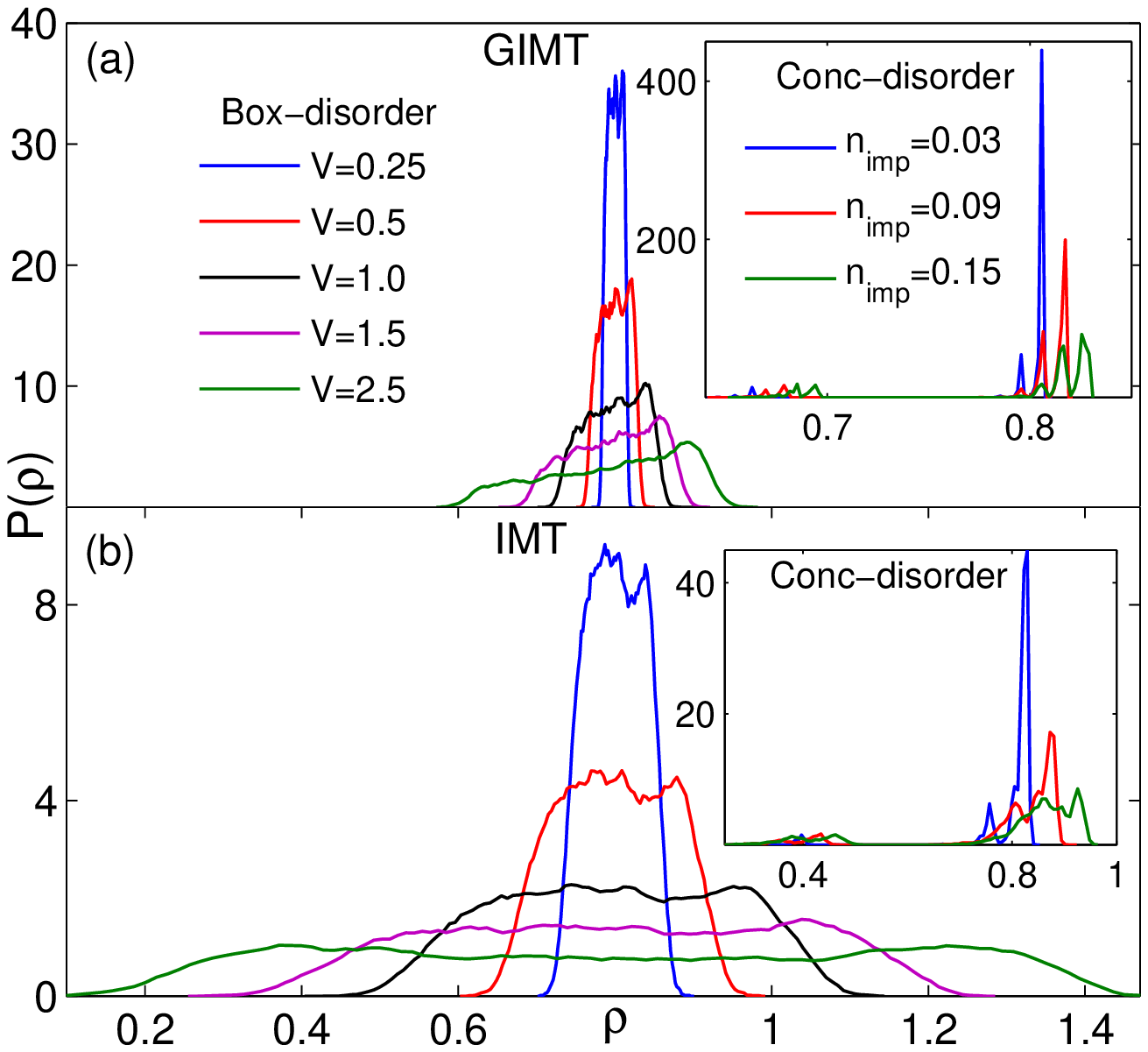}
\end{minipage}
\begin{minipage}{.9\columnwidth}
\includegraphics[width=1.0\columnwidth,clip=]{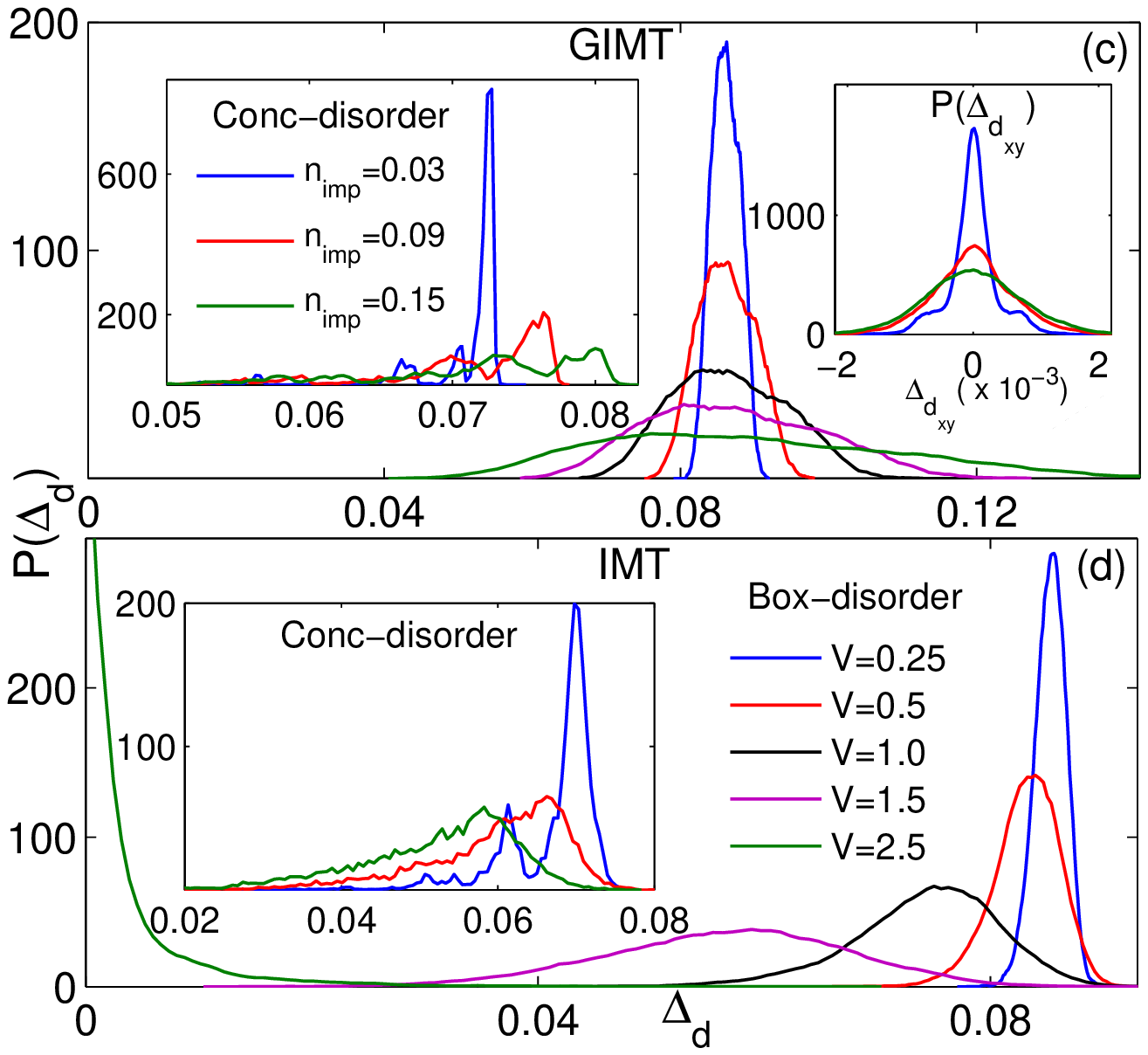}
\end{minipage}
\end{center}
\caption{
A comparison of the distribution of local density, $P(\rho)$ (a, b) and local pairing amplitude, $P(\Delta_d)$ (c, d), from the GIMT and IMT calculations with box-disorder. The results broadly indicate that both the nature and degree of inhomogeneity differ between the GIMT and IMT calculations.
(a) The results from the GIMT calculations ($\rho_i \leq 1$ for all $i$) yields a narrow $P_{\rm GIMT}(\rho)$, that widens gradually with $V$. $P_{\rm GIMT}(\rho)$ becomes progressively asymmetric with $V$ with more weight shifting to the larger values of $\rho$.
(b) The $P_{\rm IMT}(\rho)$ yields a broad distribution, particularly at large $V$, permitting a wide range of local occupation including $\rho_i > 1$.
The insets of (a, b) present $P_{\rm GIMT}(\rho)$ and $P_{\rm IMT}(\rho)$ respectively for conc-disorder with $n_{\rm imp}=0.03,0.09$ and $0.15$, and the conclusions remain similar to those from the box-disorder.
(c) The $P_{\rm GIMT}(\Delta_d)$ show a nearly Gaussian distribution which widens with $V$, but its centroid remains practically unaltered. (d) The $P_{\rm IMT}(\Delta_d)$ yields a distribution strongly peaked around $\Delta^{(0)}_d$ only for weak $V\sim 0.25$. This peak-position marches down with $V$, producing a broad hump by $V=1.5$. $P_{\rm IMT}(\Delta_d)$ becomes a skewed distribution by $V\leq 2.5$ with a large weight near $\Delta_d \approx 0$.
The left insets of (c) and (d) present the results of $P(\Delta_d)$ from the conc-disorder with $n_{\rm imp}=0.03,0.09$ and $0.15$, showing the similar qualitative features as in the main panels. The right inset in (c) shows the $P(\Delta_{d_{xy}})$ for GIMT conc-disorder calculations. The tiny width of $P(\Delta_{d_{xy}})$ compared to $P(\Delta_d)$, and its near-zero average establishes its insignificance.
}
\label{fig:Pden}
\end{figure*}

We plot along the $y$-axis of figures~\ref{fig:Pden}(a) and \ref{fig:Pden}(b) the number of times a value of $\rho$ occurs anywhere in the system against $\rho$ itself along the $x$-axis from the GIMT and IMT findings. The resulting normalized distribution $P(\rho)$ 
widens with $V$, however, the disorder dependence of $P(\rho)$ is markedly different in the two calculations. For example, at $V=2.5$, $P_{\rm GIMT}(\rho)$ is finite only for $\rho \in [0.6,0.95]$, where as $P_{\rm IMT}(\rho)$ is non-zero over a much wider range of $\rho=0.1$ to $\rho=1.45$. The Gutzwiller projection ensures $\rho_i \leq 1$.
It is interesting to note that $P_{\rm GIMT}(\rho)$ develops an asymmetry with increasing $V$ with a larger weight towards larger values of density. Such asymmetry is absent in $P_{\rm IMT}(\rho)$. The explanation for this feature, 
will be discussed in section~\ref{subsec:Spatial}. Results from the conc-disorder (which includes three-site terms and $J'$ terms in GIMT) is broadly similar to the box-disorder with respect to the inhomogeneities, and are presented as the insets in the figures~\ref{fig:Pden}(a) and \ref{fig:Pden}(b). As expected, $P(\rho)$ develops finite weight for depleted values of $\rho$ on the disorder sites (at $\rho\sim 0.7$ for GIMT and $\rho\sim 0.4$ for IMT calculations), in addition to the large contribution around its average, $\rho \approx 0.8$. It is interesting that the electron population on the impurities is much larger in GIMT than in plain IMT. This ensures that the electronic correlations prohibit strong inhomogeneities. Further, it reduces the effective disorder strength, if we were to define this strength locally in terms of depleting population of the electrons in our context of repulsive scatterers. The reduction of the effective disorder has serious consequences on the physical properties of the superconductor and will be addressed in section~\ref{subsec:EffectiveDisorder}.

The local density plays a crucial role in deciding the nature of the local d-wave order parameter. The double occupant sites (and empty sites) in the IMT calculation deplete the $d_{x^2-y^2}$-wave pairing amplitude locally, defined by $\Delta_d(i)=\frac{1}{4}(\Delta_i^{ +x} - \Delta_i^{ +y} + \Delta_i^{ -x} -\Delta_i^{ -y})$. This is because these sites have extreme values of local density (namely, $\rho_i=0$ or $2$) and do not participate in the particle-hole mixing, which is essential for a non-zero value to $\Delta_d(i)$. As a result, the d-wave pairing amplitude from plain IMT reduces with increasing $V$ (box-disorder) or with $n_{\rm imp}$ (conc-disorder). The corresponding distribution, $P_{\rm IMT}(\Delta_d)$, as shown in figure~\ref{fig:Pden}(d), broadens with increasing $V$ with large weight progressively shifting towards a smaller value. Such an evolution finally leads to a rather skewed $P_{\rm IMT}(\Delta_d)$ at large $V$, where a large number of sites have $\Delta_d(i) \approx 0$, as shown in figure~\ref{fig:Pden}(d). In fact, this same physical picture \cite{Ghosal01,Ghosal98} yielded a very similar disorder-dependence of $P_{\rm IMT}(\Delta_s)$ for a disordered sSC.
On the other hand, the d-wave pairing amplitude in the GIMT calculation, which is free of doubly-occupied sites (and hence, free of near-empty sites too, satisfying the constraint of maintaining $\rho=0.8$), stays around its homogeneous value (see figure~\ref{fig:Pden}(c)). Of course, there are inhomogeneities in the GIMT results, which give the increasing width in $P_{\rm GIMT}(\Delta_d)$ with $V$. However, the fluctuations in the pairing amplitude are weak (at least for moderate $V$) and symmetric about $\Delta^{(0)}_d$ yielding a $P_{\rm GIMT}(\Delta_d)$ nearly Gaussian in shape for all $V$.
While $P(\rho)$ indicates that the degree of inhomogeneity is different in the GIMT and IMT schemes, the results for $P(\Delta_d)$ also  establish that the nature of inhomogeneity is substantially different in the two calculations.
Such a disorder evolution of $P_{\rm GIMT}(\Delta_d)$ is strikingly different from the predictions of simple Abrikosov-Gorkov theory, or its extension in the form of self-consistent T-matrix approximation \cite{Maki01,Hirschfeld88}, in which $\Delta_d$ (assumed homogeneous) depletes monotonically with $V$.
It is interesting to note that the GIMT calculation yields $\Delta_d(i) \geq \Delta^{(0)}_d$ on certain locations, particularly at large $V$, whereas the IMT value of $\Delta_d(i)$ is more or less limited by $\Delta^{(0)}_d$. The similarity of our $P_{\rm GIMT}(\Delta_d)$ and that found from the nanoscale inhomogeneity \cite{Pan01,Lang02} in $\rm {Bi}_2\rm {Sr}_2 \rm {Ca} \rm {Cu}_2 \rm {O}_{8+x}$ (BSCCO) is heartening, while the nature of $P_{\rm IMT}(\Delta_d)$ is unfound in cuprates. 
We draw the attention of the readers to the fact that in plotting $P(\Delta_d)$ in figure~\ref{fig:Pden}, we introduced an additional prefactor $J(3g^{xy}_{ij}+1)/8$ such that the GIMT and IMT calculations produce same $\Delta_d$ at $V=0$ ($g^{xy}_{ij}=1$ for IMT). We also present the GIMT distribution of one sub-dominant Cooper-channel order for conc-disorder, $\Delta_{d_{xy}}$, as the right inset of figure~\ref{fig:Pden}(c). The tiny width of $P(\Delta_{d_{xy}})$ and also the smallness of ${\rm max}\{\Delta_{d_{xy}}(i)\}$ at all $V$ illustrate the insignificance of $\Delta_{d_{xy}}$ compared to $\Delta_d$. Their redundancy is further verified by repeating the GIMT calculations where $\Delta_{d_{xy}}$, $\Delta_{s_{xy}}$ and $\Delta_{s_{xs}}$ are suppressed by hand, without showing any change in the results. The disorder dependence of $P(\Delta_{s_{xy}})$ and $P(\Delta_{s_{xs}})$ are essentially same as that of $P(\Delta_{d_{xy}})$. Upon establishing their insignificance we drop all the sub-dominant orders for the box-disorder calculation in favor of simplicity and clarity. 

\subsection{Disorder dependences of observables}\label{subsec:PhysObs}

\begin{figure*}
\begin{center}
\begin{minipage}{.87\columnwidth}
\includegraphics[width=1.0\columnwidth,clip=]{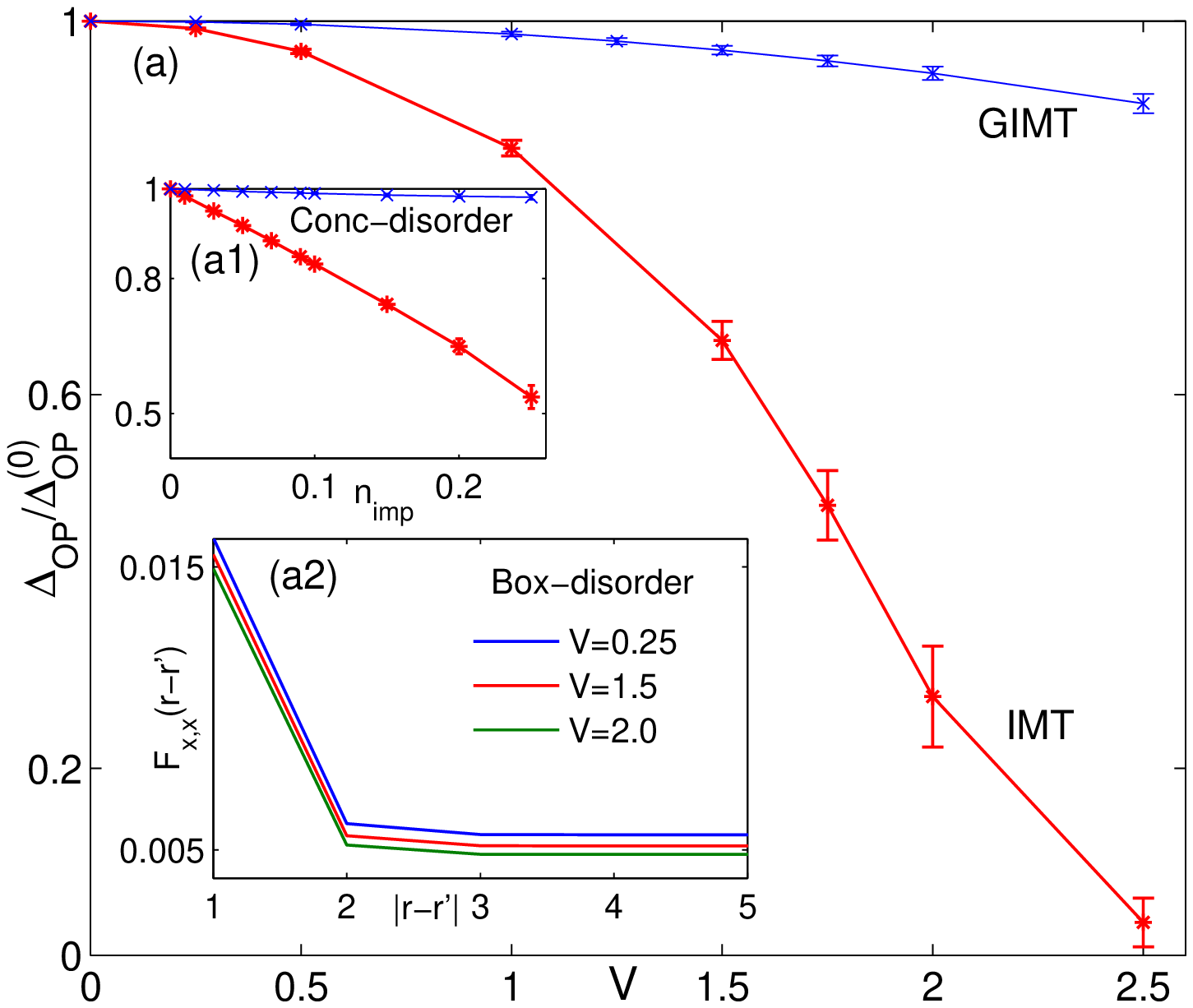}
\end{minipage}
\begin{minipage}{.93\columnwidth}
\includegraphics[width=1.0\columnwidth,clip=]{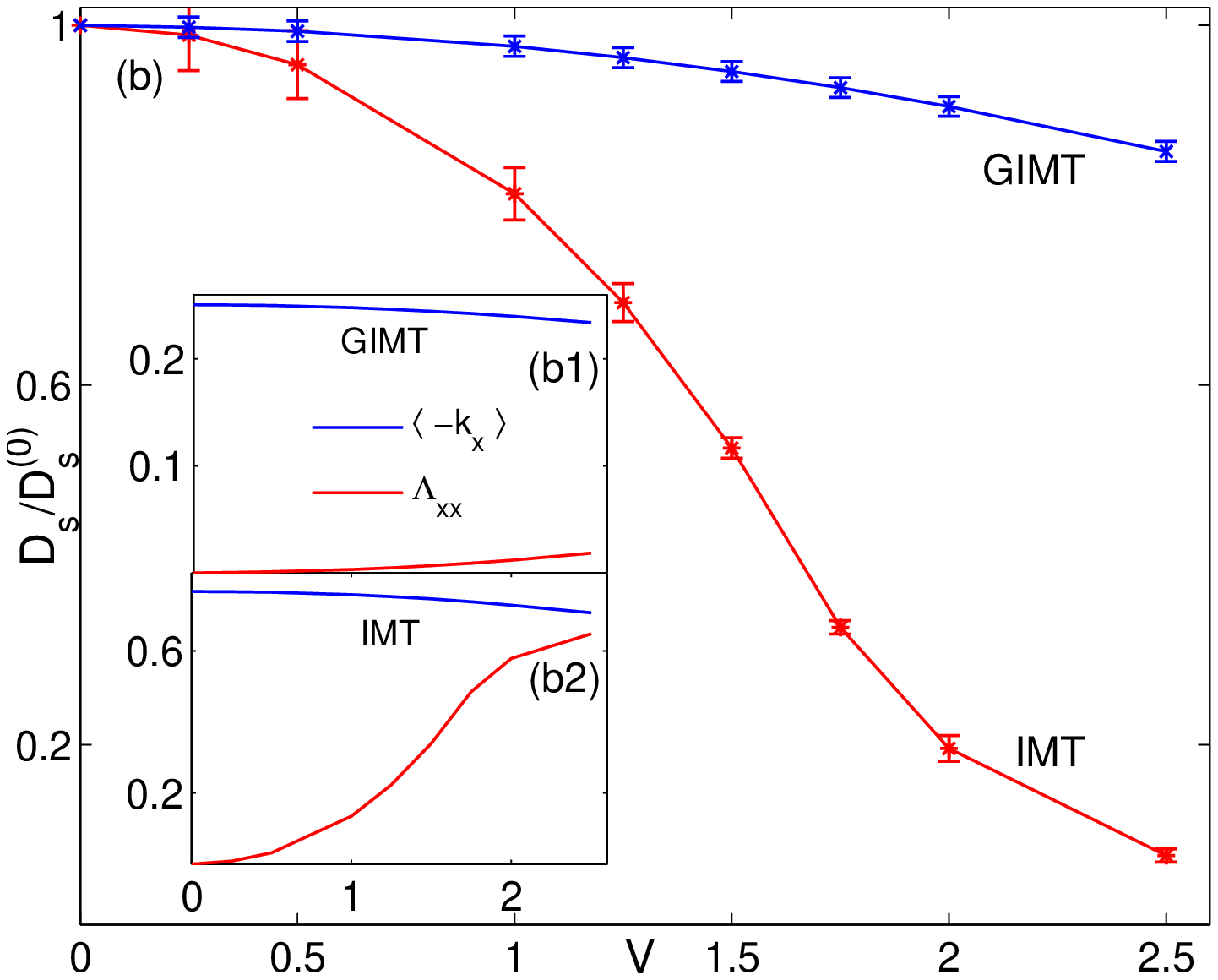}
\end{minipage}
\end{center}
\caption{Variation of disorder averaged $\Delta_{\rm OP}$ (a) and $D_s$ (b), (normalized by their respective values at $V=0$) with $V$ (box-disorder). The GIMT results are given by the blue lines, showing a very weak $V$-dependence of $\Delta_{\rm OP}$ and $D_s$. The IMT results for $\Delta_{\rm OP}$ and $D_s$ (red lines) drop severely with $V$, both collapsing by $V \approx 3$, marking a transition to a non-superconducting state. (a) Such a weak variation of $\Delta_{\rm OP}$ in GIMT is consistent with the $P(\Delta_d)$ in figure~\ref{fig:Pden}(c), once we identify $\Delta_{\rm OP} \approx \langle \Delta_d(i) \rangle$ (see \ref{sec:appendix2a}). The inset (a1) presents a qualitatively similar trend of $\Delta_{\rm OP}$ from conc-disorder, though its reduction in the IMT (red line) is not as dramatic. The inset (a2) shows the insensitivity of the length-scale of decay of $F_{\delta, \delta'}(i-j)$ (defined in text) on $V$. (b) Errors in $D_s$ arise largely from the extrapolation of $\Lambda_{xx}(q_y \rightarrow 0)$ from a finite-size simulation, becoming less accurate for small $V$. Insets: Evolutions of $\langle -k_x \rangle$ and $\Lambda_{xx}(q_y \rightarrow 0)$ from the GIMT and IMT calculations in (b1) and (b2) respectively. The results show that only a tiny fraction of $\Lambda_{xx}(q_y \rightarrow 0)$ arises from $V$ in the GIMT with respect to plain IMT contributions. The $V$ dependence of $\langle -k_x \rangle$ is quite similar from the two methods, though its bare value is suppressed in the GIMT calculation by the local  $g^{t}_{ij}$.
}
\label{fig:odlro}
\end{figure*}

The qualitative as well as quantitative differences in the GIMT and plain IMT results for $P(\rho)$ and $P(\Delta_d)$ raises the question: How do the inhomogeneities, or the lack of it in the presence of strong correlations, affect the disorder dependence of various physical observables? We address this issue in figures~\ref{fig:odlro} and \ref{fig:dos} where we show the evolution of ODLRO, superfluid density and average DOS with $V$. The results bring out a striking contrast in the $V$-dependence of these quantities with or without the strong correlations. 

\subsubsection{Off-diagonal long range order (ODLRO)}\label{subsec:ODLRO}

First, we discuss the superconducting ODLRO defined as: 
\begin{equation}
\Delta_{\rm OP}^2=\lim_{|i-j| \to \infty} F_{\delta, \delta'}(i-j)
\label{eq:odlro}
\end{equation}
where $F_{\delta, \delta'}(i-j)=\langle B_{i \delta}^{\dagger} B_{j \delta'} \rangle$, and, $B_{i \delta}^{\dagger}=( c^{\dagger}_{i \uparrow} c^{\dagger}_{i + \delta \downarrow} + c^{\dagger}_{i + \delta \uparrow} c^{\dagger}_{i  \downarrow})$ is the singlet Cooper-pair creation operator on the links connecting the neighboring sites at $i$ and $i+\delta$. $F_{\delta, \delta'}(i-j)$ is independent of $\delta$ and $\delta'$ in the limit of large $|i-j|$. Since $F_{\delta, \delta'}(i-j)$ can be interpreted as simultaneous hopping of a singlet cooper-pair on a link, the GRF corresponding to this process is $g^{t}_{i,j}g^{t}_{i+\delta,j+\delta'}$.
The evolution of $\Delta_{\rm OP}$ (normalized by its value $\Delta^{(0)}_{\rm OP}$ at $V=0$) with box-disorder is presented in figure~\ref{fig:odlro}(a). With increasing $V$, $\Delta_{\rm OP}$ from the IMT results decreases monotonically, initially with a gradual fall, followed by a rapid decrease beyond $V \ge 1$, causing it to almost vanish by $V \sim 2.5$. Such a behavior is consistent with the $V$-dependence of $P_{\rm IMT}(\Delta_d)$ in figure~\ref{fig:Pden}(d). The strong correlations make the superconductor robust against disorder. There is hardly any significant fall of the GIMT value of $\Delta_{\rm OP}$ at $V=2.5$, a strength by which the transition to a non-superconducting state occurs within the IMT scheme. The Gaussian-type shape of $P_{\rm GIMT}(\Delta_d)$ with minimally depleted centroid for all $V$ is crucial to ascertain that $\Delta_{\rm OP}$ sees little changes in the whole range of $V$ within the GIMT calculation.
The comparison of the GIMT and IMT results for $\Delta_{\rm OP}$ from conc-disorder, as presented in the inset of figure~\ref{fig:odlro}(a), shows a trend similar to the box-disorder results. A numerically inexpensive scheme for approximating $\Delta_{\rm OP}$ is further discussed in \ref{sec:appendix2a}.

The length scale of decay of $F_{\delta, \delta'}(i-j)$ with $|i-j|$ defines the coherence length ($\xi$) of a superconductor. The evolution of $F_{\delta, \delta'}(i-j)$ for several $V$, as presented in the lower inset of figure~\ref{fig:odlro}(a) from the GIMT method, implies that there is hardly any $V$-dependence of $\xi$. 
Given that $\xi$ is already small (only a few lattice spacings) in the clean cuprates, it is not expected to show any significant reduction with disorder. A similar insensitivity of $\xi$ on $V$ is also found in the IMT results.

\subsubsection{Superfluid stiffness}\label{subsec:DS}

The defining characteristic of a superconductor lies in the Meissner \cite{Tinkhambook04} effect, which is quantified by the stiffness of the BCS-GS wave function to an externally applied phase twist. This stiffness translates into its finite superfluid stiffness, which is proportional to the superfluid density. Within the Kubo formalism the superfluid stiffness, denoted as $D_s$, is measured by,
\begin{equation}
\frac{D_s}{\pi}=\langle -k_x \rangle - \Lambda_{xx}(q_x=0,q_y \rightarrow 0,\omega=0),
\label{eq:supd}
\end{equation}
where $k_x$ is the kinetic energy along the $x$-direction and $\Lambda_{xx}$ is the long wavelength limit of transverse (static) current-current correlation function \cite{Scalapino93}. It is calculated by Fourier transforming the impurity averaged Matsubara Green's function;
\begin{equation}
\Lambda_{xx}(\mathbf{q},i\omega_n)=\frac{1}{N}\int_0^{1/T}d\tau e^{i\omega_n\tau}\langle j_x^p(\mathbf{q},\tau)j_x^p(-\mathbf{q},0)\rangle,
\label{eq:lambda}
\end{equation}
where $j_x^p(\mathbf{q})$ is the paramagnetic current and $\omega_n=2\pi nT$ ($n$ is a positive integer). The GRF for $\langle k_x\rangle$ is $g^{t}_{i,i+x}$ and it is $g^{t}_{i,i+x}g^{t}_{j,j+x}$ for $\Lambda_{xx}(\mathbf{r_i},\mathbf{r_j},\tau)$. In order to obtain a good $\mathbf{q}$-resolution of $\Lambda_{xx}$ from a finite simulation box, we used an effectively larger system employing `repeated zone scheme' (RZS) (see \ref{sec:appendix3}) for a single realization of disorder.
The final $q_y \rightarrow 0$ extrapolation of the disorder averaged $\Lambda_{xx}$ obtained from single unit cell was guided by $\Lambda^{\rm RZS}_{xx}(q_y)$. The disorder dependence of $\Delta_{\rm OP}$ and $D_s$ are quite similar, while the IMT causes them to almost vanish by $V \sim 2.5$, there is hardly any appreciable reduction of them in the GIMT results.

The inset of figure~\ref{fig:odlro}(b) shows the individual disorder dependence of $\langle -k_x \rangle$ and $\Lambda_{xx}(q_y \rightarrow 0)$. The $\langle -k_x \rangle$ from the GIMT method are smaller than the IMT values due to a complete removal of double-occupancy. The kinetic energy, which is the diamagnetic contribution to $D_s$, decreases slowly with disorder and the rate of decrease is similar for both the IMT and GIMT calculations, because it is proportional to the total density. For $V=0$, $\Lambda_{xx}$ is zero in the absence of any quantum fluctuations making $D_s$ equal to $\langle -k_x \rangle$. With increasing $V$, $\Lambda_{xx}$ grows in the IMT scheme and almost equals $\langle -k_x \rangle$ by $V \sim 2.5$ and continue to reduce their differences for larger $V$. However, $\Lambda_{xx}$ remains insensitive to $V$ within the GIMT scheme making $D_s$ robust to impurities.

Such a strong insensitivity of $D_s$ to $V$ is surprising though consistent with the broad experimental trends \cite{Nachumi96}. 
There are recent data on the temperature and doping dependences of $D_s$ \cite{Hetel11} for the cuprates. We hope similar experiments will be extended to careful disorder dependence for a justified comparison with our results.

\subsubsection{Density of states (DOS)}\label{subsec:DOS}
 
\begin{figure*}
\begin{center}
\begin{minipage}{.91\columnwidth}
\includegraphics[width=1.0\columnwidth,clip=]{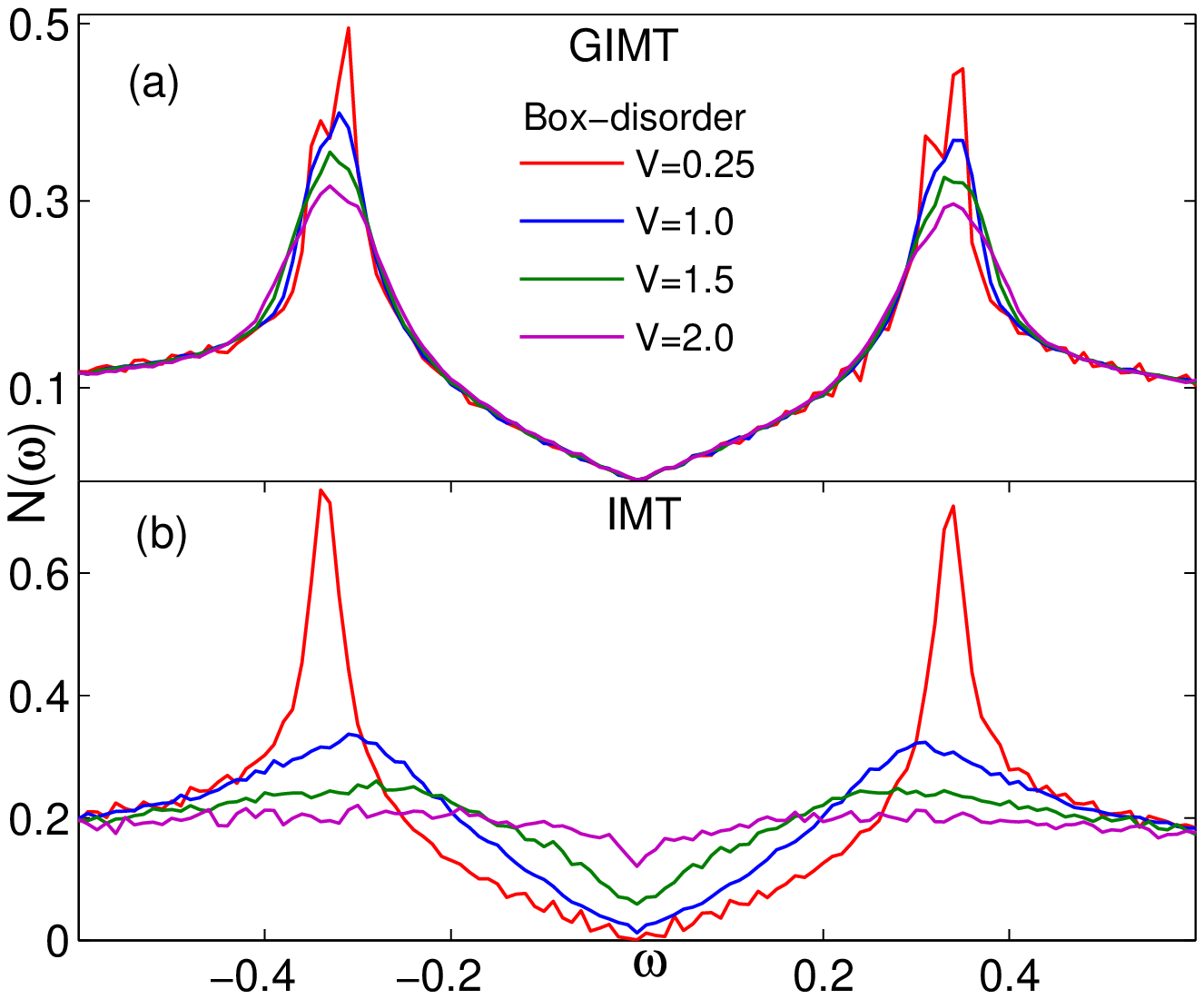}
\end{minipage}
\begin{minipage}{.89\columnwidth}
\includegraphics[width=1.0\columnwidth,clip=]{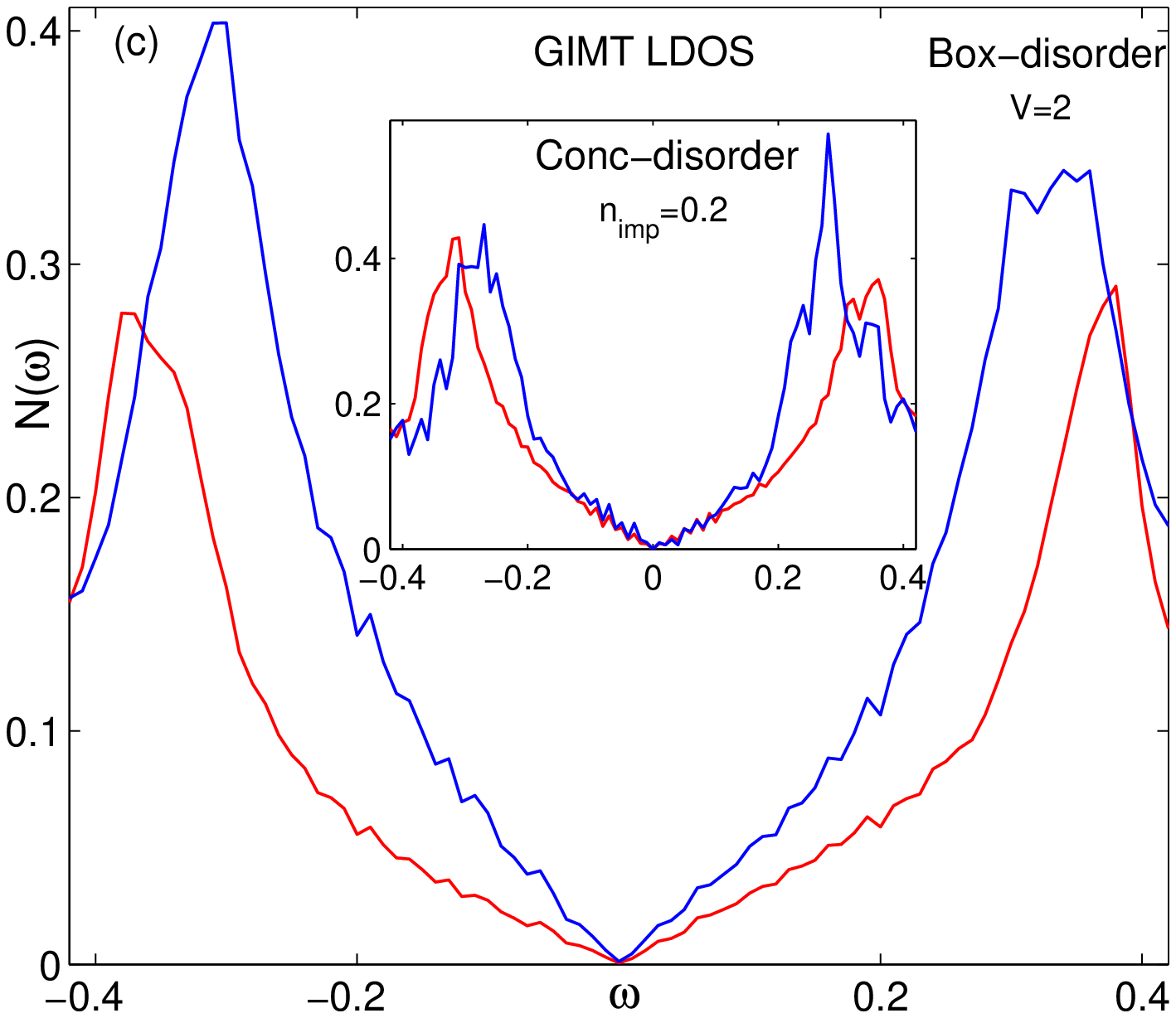}
\end{minipage}
\end{center}
\caption{
(a) Traces of site-averaged DOS, $N_{\rm GIMT}(\omega)$ for different $V$ (box-disorder) stay exactly on top of one another leaving the linear V-shape low-lying DOS of a pure dSC unaltered.
The coherence peaks suffer a weak reduction of weight
with $V$. However, their positions hardly change from $\pm \Delta^{(0)}_d$. (b) The $N_{\rm IMT}(\omega)$ undergoes gap-filling for small $|\omega|$ with increasing $V$, in sharp contrast to the GIMT findings. The IMT coherence peaks collapse with $V$ and their locations move to a lower $|\omega|$ with $V$. (c) The local density of state (LDOS) from the GIMT calculations at $V=2$ (box-disorder) on regions of {\it locally} large and small $\Delta_d$. The LDOS in the two regions are identified by its average over all sites with large $\Delta_d(i) \in [0.25,0.28]$ (shown by the red trace) and with small $\Delta_d(i) \in [0.18,0.21]$ (shown by the blue trace). While the anticorrelation in the coherence peak-height and its location is clear, the GIMT calculations show different low-$\omega$ behavior for $N(\omega)$ between the large and small $\Delta_d$ regions.
A homogeneity of low-lying LDOS, however, is more pronounced up to $\omega \sim \Delta^{(0)}_{d}/2$ in case of the conc-disorder, as shown in the inset.
}
\label{fig:dos}
\end{figure*}

We now turn towards the discussion of the disorder-dependence of the site-averaged DOS, given by:
\begin{equation}
N(\omega)=\frac{1}{N}\sum_{i,n} g^t_{ii} \left\{ |u_{i,n}|^2\delta(\omega-E_n) + |v_{i,n}|^2\delta(\omega+E_n) \right\}
\label{eq:dos}
\end{equation}
where $E_n$'s are the BdG eigenvalues. We focus on $N(\omega)$ for $\omega$ up to the coherence peak energy. While $N(\omega)$ consists only of $\delta$-function peaks at $T=0$, it is necessary to obtain a smooth trace such that the relevant features are truly identified and is free of any artefact of smoothing.
In order to achieve this, we generate a denser spectrum by using RZS (see \ref{sec:appendix3}). We broaden each of the $\delta$-function in $N(\omega)$ by an amount comparable to the average spacing of the $E_n$'s, and the final $N(\omega)$ is presented in figures~\ref{fig:dos}(a) and \ref{fig:dos}(b). After establishing that the additional Cooper-channel orders, e.g., $\Delta_{s_{xs}}$, $\Delta_{s_{xy}}$, $\Delta_{d_{xy}}$ have insignificant contribution to $N(\omega)$, just as in all the other observables, we restrict the DOS calculations implemented with RZS only to $t-t'-J$ model for all classes of disorder.

While the average $N(\omega)$ calculated within plain IMT shows significant gap filling in figure~\ref{fig:dos}(b) for large $V$, results from the GIMT scheme in figure~\ref{fig:dos}(a) show that the low-lying DOS remains unaffected with impurities even up to $V=3$. Strong correlations thus appear to prohibit the formation of the low-energy states in the GIMT results. While the coherence peak height gets reduced with $V$ in both the methods, the rate of this reduction is significantly weaker in the GIMT. Our results for conc-disorder are similar to Ref. \onlinecite{Garg08}. The strong correlations thus seem to protect the V-shape of the $N(|\omega| \leq \Delta^{(0)}_d)$, a feature established in the STS-experiments \cite{McElroy05,Jenny02}.

In order to develop a deeper understanding of the robustness of low-energy DOS in GIMT, we calculated the LDOS for two regions with large and small $\Delta_d$. We average the LDOS for these regions with a weak $\Delta_d(i)  \in [0.18,0.21]$ and with a strong $\Delta_d(i)  \in [0.25,0.28]$ in figure~\ref{fig:dos}(c) and find that LDOS in the two regions behave quite differently even for $N(|\omega| \leq \Delta^{(0)}_d)$. We will also show in section~\ref{subsec:Spatial} that the regions of large (small) $\Delta_d(i)$ corresponds also to the regions of large (small) $\rho_i$ within the GIMT calculations, so that the inhomogeneity in the local density is also reflected in the same LDOS. While the anti-correlation between the coherence peak heights and the local gap values is clearly identified similar to experiments \cite{McElroy05}, we did not find the expected homogeneity in LDOS for $|\omega|<\Delta^{(0)}_d$. Such homogeneity is recovered in the site-averaged $N(\omega)$, which can be easily seen from the $P(\Delta_d)$ in figure~\ref{fig:Pden}(c). The peak remains symmetric for all $V$ centered around $\Delta^{(0)}_d$. For the conc-disorder, the homogeneity of the low-lying LDOS is maintained fairly well \cite{Garg08} up to $\omega \sim \Delta^{(0)}_{d}/2$. This is shown in the inset presenting the LDOS for sites with $\Delta_d(i) \in [0.14,0.17]$ (blue curve) and for sites with $\Delta_d(i) \in [0.19,0.22]$ (red curve).

\begin{figure*}[t]
\centering
\includegraphics[angle=0, width=0.65\textwidth]{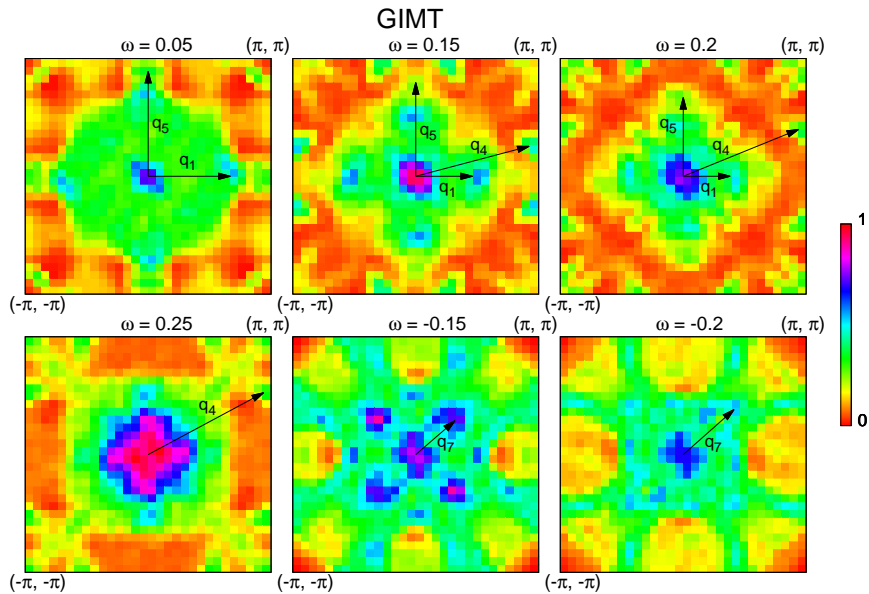}
\caption{
The colour-density plot of $N(q,\omega)$ for $n_{\rm imp}=0.09$ (conc-disorder) from the GIMT scheme. Results are presented in the first Brillouin zone for different values of $\omega$. The intensity of the peak at $\mathbf{q}=0$ has been truncated for the clarity. We used the scaling, $x \rightarrow (x - x_{\rm min})/(x_{\rm max} - x_{\rm min})$, such that $x \in [0:1]$ for all panels (here $x=N(q,\omega)$). The truncation value, $x_{\rm max}=1,4,8,11,5~{\rm and}~10$ (arbitrary units) for $\omega=0.05,0.15,0.2,0.25,-0.15~{\rm and}~-0.2$ respectively. Corresponding (linear) colour scale is given at the right of the figure. Several octet-peaks (indicated by arrows) can be recognized, all of which disperse with $\omega$. The asymmetry between $N(q,\pm \omega)$ is also evident.
}
\label{fig:ftldos}
\end{figure*}

\subsubsection{Fourier transformed local density of states (FT-LDOS)}\label{subsec:FTLDOS}

Another observable, that received attention in the recent times, is the energy resolved FT-LDOS, defined by $N(\mathbf{q},\omega)=\sum_i e^{-i\mathbf{q}.\mathbf{r}_i} N_i(\omega)$,
where $N_i(\omega)$ is the LDOS at site $i$, obtained from equation~(\ref{eq:dos}) without performing the summation over $i$. FT-LDOS is extracted from the STS data indicating the existence of a periodic and energy resolved modulation of the local density.
An interpretation of the STS data comes from the simple `octet'-model \cite{McElroy03,Wang03}, based on the scattering of the d-wave quasiparticles from a single impurity. The resulting theory yields an octet of wave-vectors $\mathbf{q}$ connecting the constant quasiparticle-energy contours for which the scattering probability is maximal.
The Fermi surface obtained from the dispersion of these octet-members, agrees well with its direct measurement from the angle-resolved photo-emission spectroscopy \cite{Shen08}. The extension of the octet-analysis to many-impurities is a subtle issue. Without invoking strong correlations, results from many-impurity calculations found sensitive dependence on the details of the disorder \cite{Nunner06} for a reasonable fit to the STS data, and the octet peaks are often masked by a large noise \cite{Zhu04}. Further, non-dispersive FT-LDOS peaks have been attributed to competing orders \cite{Ghosal05} for under-doped samples. However, to the best of our knowledge, the role of strong correlations on the FT-LDOS have never been addressed in a calculation.

We present the FT-LDOS from the GIMT calculations in figure~\ref{fig:ftldos}.
Each density-plot shows the power spectrum $P(\mathbf{q},\omega)=\overline{|N(\mathbf{q},\omega)|^2/N}$ for a given $\omega$ on the first Brillouin zone (BZ): $q_x,q_y \in [-\pi,\pi]$, disorder averaged over $10$ independent realizations of conc-disorder with $n_{\rm imp}=0.09$. We apply RZS for a smooth data from a larger system (see \ref{sec:appendix3}). Octet peaks can easily be recognized, such as $\mathbf{q}_1 \sim (0,\pm\pi/{\cal L}_1)$ or $(\pm\pi/{\cal L}_1,0)$. The dispersion of this peak results into a continuous change in ${\cal L}_1\sim 1.36$ for $\omega=0.05$ to ${\cal L}_1\sim 7.48$ for $\omega=0.2$.
Another (dispersing) peak, likely the $\mathbf{q}_7$ in the octet terminology \cite{McElroy03}, can also be discerned along the diagonal direction of the BZ (better resolved for negative $\omega$).
Our results recover much of the experimental trends, e.g., the shape and location of the FT-LDOS peaks, their dispersion (except may be $\mathbf{q}_5$), among other things. However, the asymmetry of the FT-LDOS profiles for $\pm\omega$ is found to be much stronger than the experimental data.
The qualitative features of figure~\ref{fig:ftldos} remain unaltered with a change of $n_{\rm imp}$ to $0.05$ from $0.09$. A similar calculation of FT-LDOS within the IMT scheme (shown in the supplementary material) was found to capture the key points of the figure~\ref{fig:ftldos}, however, appropriate choice of the model-parameters for the IMT calculations is crucial for a justified comparison.

Before closing the discussions on the disorder dependence of the observables, we note that the degree of inhomogeneity in $\Delta_d$ reduces upon including strong electronic repulsions (See figure~\ref{fig:Pden}(c)), at least for the moderate disorder ($0.5 \leq V \leq 2.0$). However, the behavior of the observables remain very different compared to the self-consistent T-matrix predictions \cite{Maki01} based on the assumption of homogeneous but renormalized $\Delta_d$ for any $V$. The inadequacy of self-consistent T-matrix approximation for describing disordered cuprates is discussed in Ref. \onlinecite{Hirschfeld00}, which is a weak coupling theory built on simple metallic normal state. The strong interactions amplify such differences further due to modifications of the normal states, as we will see in section~\ref{subsec:Normal}. 

\subsection{Spatial distribution of local orders and their relation to inhomogeneity}\label{subsec:Spatial}

Having seen the dramatic robustness of the observables to impurities in the GIMT scheme, it appears natural to look for the relationship between such insensitivity and the inhomogeneity compared to plain IMT. We develop an important insight by studying the spatial distribution of the different order parameters on the lattice, as we discuss below.

\begin{figure*}
\begin{center}
\begin{minipage}{.9\columnwidth}
\includegraphics[width=1.0\columnwidth,clip=]{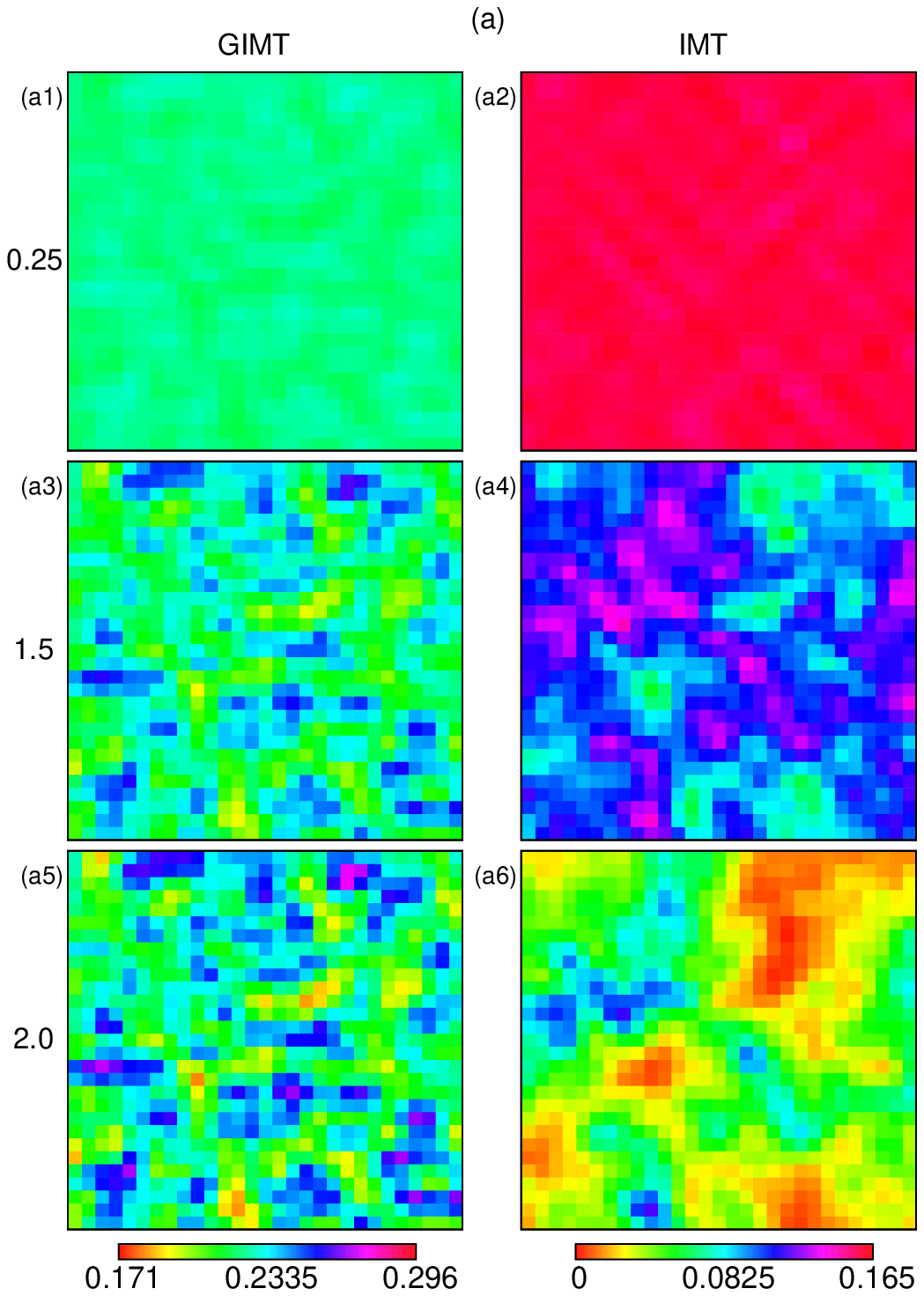}
\end{minipage}
\begin{minipage}{.9\columnwidth}
\includegraphics[width=1.0\columnwidth,clip=]{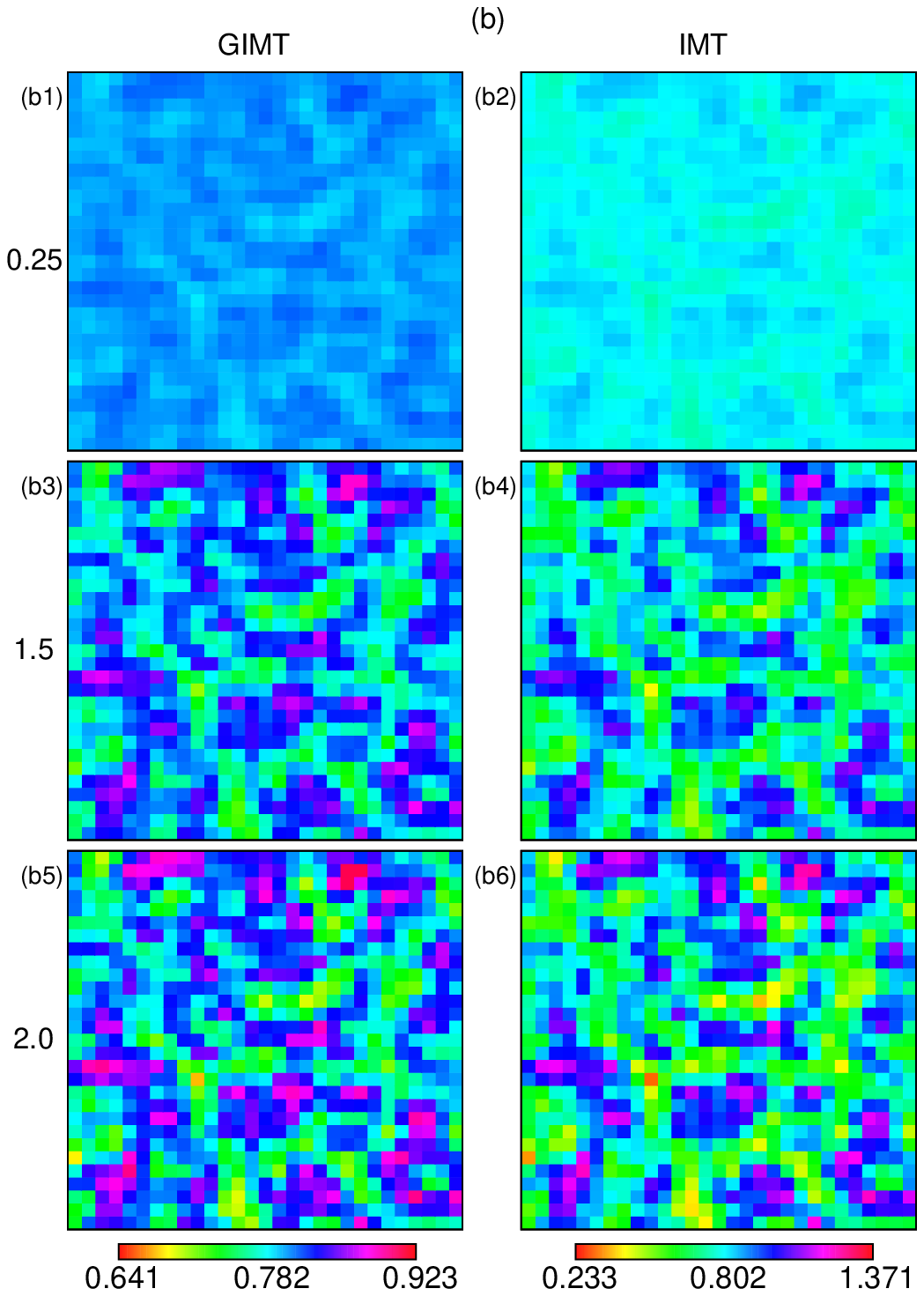}
\end{minipage}
\end{center}
\caption{
The spatial profile of (a) $\Delta_d(i)$, and (b) $\rho_i$ on a $30\times 30$ lattice for one realization of box-disorder with $V=0.25$ (a1,a2,b1,b2), $V=1.5$ (a3,a4,b3,b4) and $V=2.0$ (a5,a6,b5,b6). The left and right columns of (a) and (b) show results from the GIMT and IMT calculations respectively, and the corresponding (linear) colour-scale is given below each column. The results illustrate that the inhomogeneities in both $\Delta_d(i)$ and $\rho_i$ are weaker in the GIMT than in the IMT results (note the difference in the colour-scale between the two). A comparison between (a) and (b) shows that both the $\rho_i$ and $\Delta_d(i)$ from GIMT are large or small approximately in the same spatial locations, whereas the two differ badly in the IMT scheme. (a) The spatial inhomogeneities of the GIMT $\Delta_d(i)$ do not essentially evolve beyond $V \approx 1$. On the contrary, plain IMT results show a progressive spatial conglomeration with $V$ of sites with large $\Delta_d$ separated by regions where $\Delta_d(i)\approx 0$, leading to the formation of $\Delta_d$-`puddles'. (b) Comparison of $\rho_i$ from the GIMT and IMT profiles illustrates that the GIMT supports more sites with $\rho_i > \rho$ compared to the IMT findings. Detailed understanding of this asymmetry follows from figures~\ref{fig:veffdis}(a) and \ref{fig:Vscatter}(a).
}
\label{fig:spatialstruc}
\end{figure*}

We present the GIMT and IMT results for $\Delta_d(i)$ on the left and right columns of figure~\ref{fig:spatialstruc}(a) respectively. Similarly, the local density, $\rho_i$, from the GIMT and IMT calculations is shown in the left and right columns of figure~\ref{fig:spatialstruc}(b). All results are shown for a particular realization of box-disorder. 
There are important points to note here: Firstly, the spatial inhomogeneity in density from the GIMT calculation does not evolve with $V$ (not even in its magnitude) as seen 
from figures~\ref{fig:spatialstruc}(b1), (b3) and (b5). While this observation reinforces the finding of figure~\ref{fig:Pden}(a), the additional information here is on the spatial structures, as explained below. The profiles of $\rho_i$ in plain IMT, shown in 
figures~\ref{fig:spatialstruc}(b2), (b4) and (b6), are similar to that from the GIMT
in figures~\ref{fig:spatialstruc}(b1), (b3) and (b5) respectively, though its magnitude differs significantly (See also the figure~\ref{fig:Pden}(b)). Secondly, the evolution of spatial profile in $\Delta_d(i)$ from GIMT with $V$ is also very weak as shown
in figures~\ref{fig:spatialstruc}(a1), (a3) and (a5), compared to the same calculated in plain IMT scheme as given
in figure~\ref{fig:spatialstruc}(a2), (a4) and (a6). Thirdly, an impressive spatial correlation is found between the GIMT profiles of $\rho_i$ and $\Delta_d(i)$ for all $V$ implying that both are large (or small) approximately on the similar region of space (further illustrated in figure~\ref{fig:Vscatter}(b)), whereas such spatial correlations are completely absent in the IMT results. The IMT data in figure~\ref {fig:spatialstruc}(a6) illustrates the formation of `islands' with large $\Delta_d(i)$, separated by regions where $\Delta_d(i) \approx 0$. We verify that the sites with large $\Delta_d$ are indeed the sites where $|V_i-\mu^{\rm HS}_i|\approx 0$ for large $V$, and correspond to the spatial regions allowing maximum particle-hole mixing in plain IMT.

Clearly, the origin of the inhomogeneities leading to the island formation in $\Delta_d(i)$ within the IMT scheme is essentially the same as in the case of an sSC \cite{Ghosal01,Ghosal98}. The lack of the evolution of the GIMT spatial structures in $\Delta_d(i)$, on the other hand, illustrates that there are no well defined `islands'.
Primarily, those sites lying in the deepest valleys of the disorder potential become nearly doubly occupied in plain IMT. As a result, no significant spatial correlation between $\Delta_d(i)$ and $\rho_i$ are expected at large $V$, which is clear from our data. Strong electronic correlations in the GIMT, however, ensure that the local density never goes beyond half-filling anywhere in the system resulting into the observation that the sites with large $\rho_i$ (close to half filling) also satisfy $|V_i-\mu^{\rm HS}_i|\approx 0$ approximately (at large $V$) for $\rho=0.8$. It is thus not surprising that we see a strong spatial correlation between local density and pairing amplitude in the GIMT results on the left columns of figures~\ref{fig:spatialstruc}(a) and \ref{fig:spatialstruc}(b) (See also figures~\ref{fig:Vscatter}(b1) and (b2)), quite in contrast with the IMT findings.
This is further illustrated in the supplementary material for out-of-plane disorder where the contrast is even more prominent.

Above observation raises a conceptual question: Inhomogeneities in $\Delta_d(i)$ within the IMT calculations arise with a length-scale of $\xi$, while the fluctuations in $\rho$ follows (uncorrelated) disorder with the natural length-scale $k^{-1}_F$, the Fermi wavelength. This makes a clear separation of the two length-scales. But, such a distinction is difficult as the spatial profile of one follows the other in the GIMT results! This turns out not so crucial for the cuprates, in which both these scales are of the order of a few lattice spacing. In fact, a recent extraction \cite{Alessandro13} of the length scales associated with disorder seems to validate the GIMT picture.

\begin{figure*}
\includegraphics[angle=0, width=0.7\textwidth]{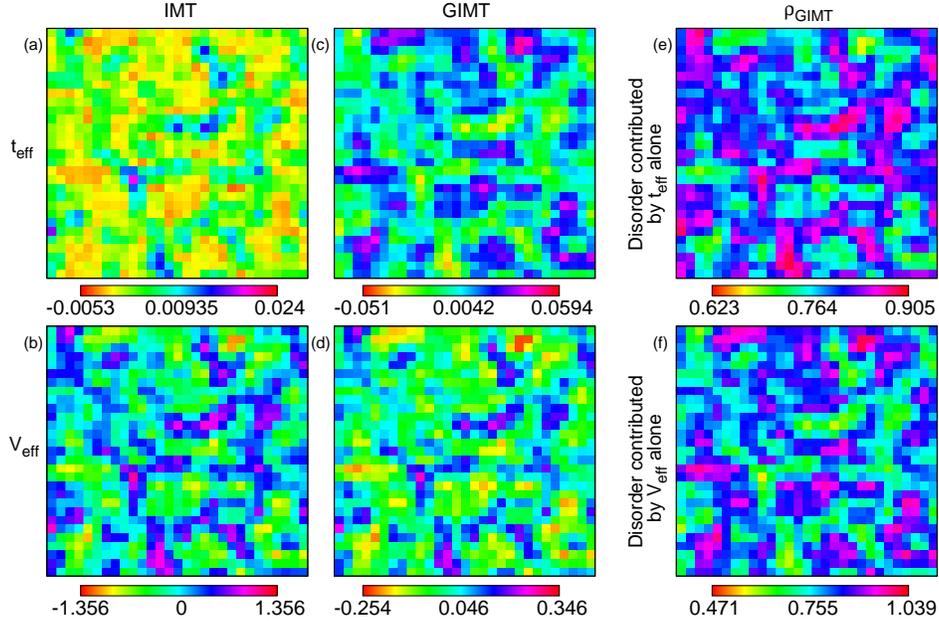}
\caption{
The spatial plots of $t_{\rm eff}$ and $V_{\rm eff}$ for one realization ($V=2$, box-disorder) are presented in 
(a) and (b) respectively from the IMT calculations. The colour-scale shows the approximate spatial correlations in the two with respect to the regions with large and small values. The magnitude show that the $t_{\rm eff}$ is more or less homogeneous in space.
Similar colour-density plots from the GIMT calculations in (c) and (d)
show a strong spatial anti-correlation between the $t_{\rm eff}$ and the $V_{\rm eff}$, see sect.~\ref{subsec:EffectiveDisorder} for further details.
The GIMT profiles of $\rho_i$ are presented in (e) and (f) where the sole source of disorder is $t_{\rm eff}$ and $V_{\rm eff}$ respectively.
They illustrate the opposite responses by these competing components of effective disorder.
}
\label{fig:tV}
\end{figure*}

\begin{figure*}[t]
\begin{center}
\begin{minipage}{.84\columnwidth}
\includegraphics[width=1.0\columnwidth,clip=]{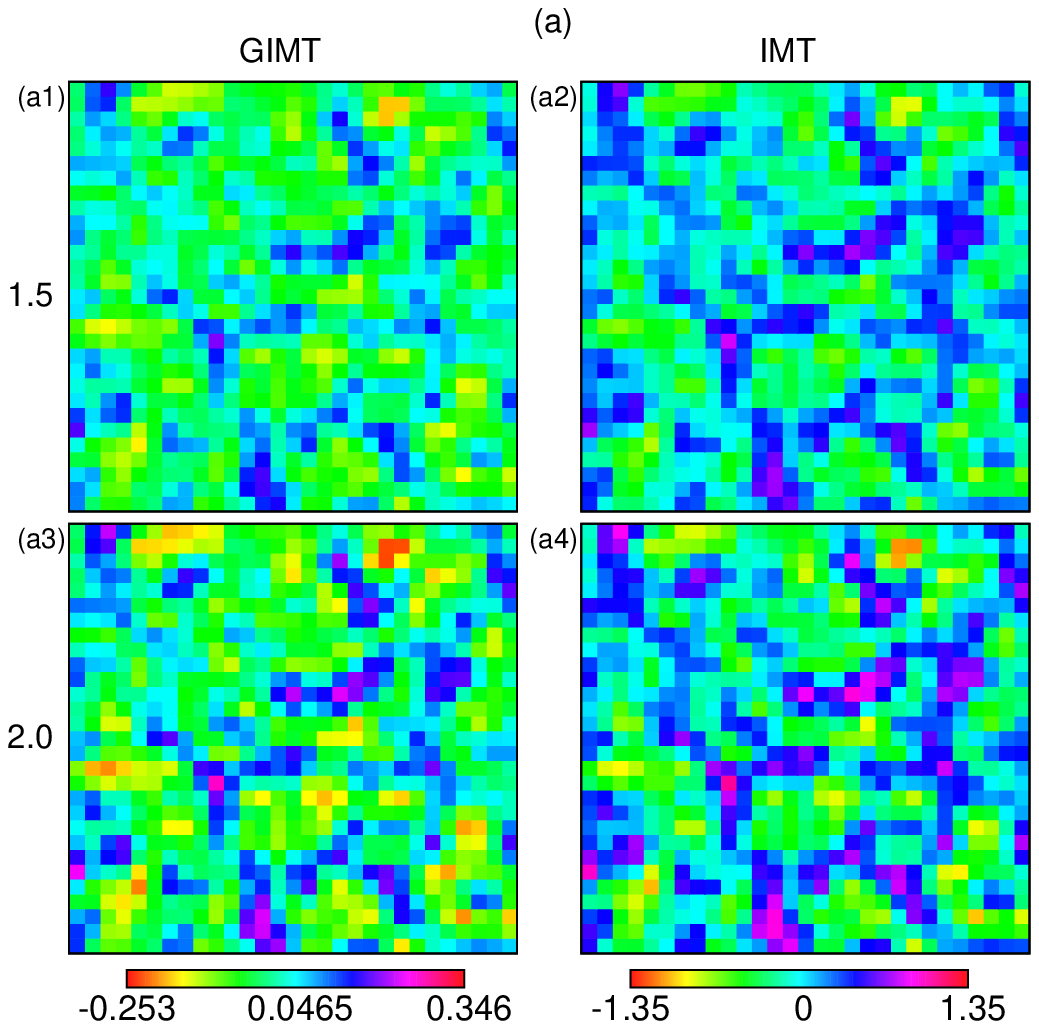}
\end{minipage}
\begin{minipage}{1.0\columnwidth}
\includegraphics[width=1.0\columnwidth,clip=]{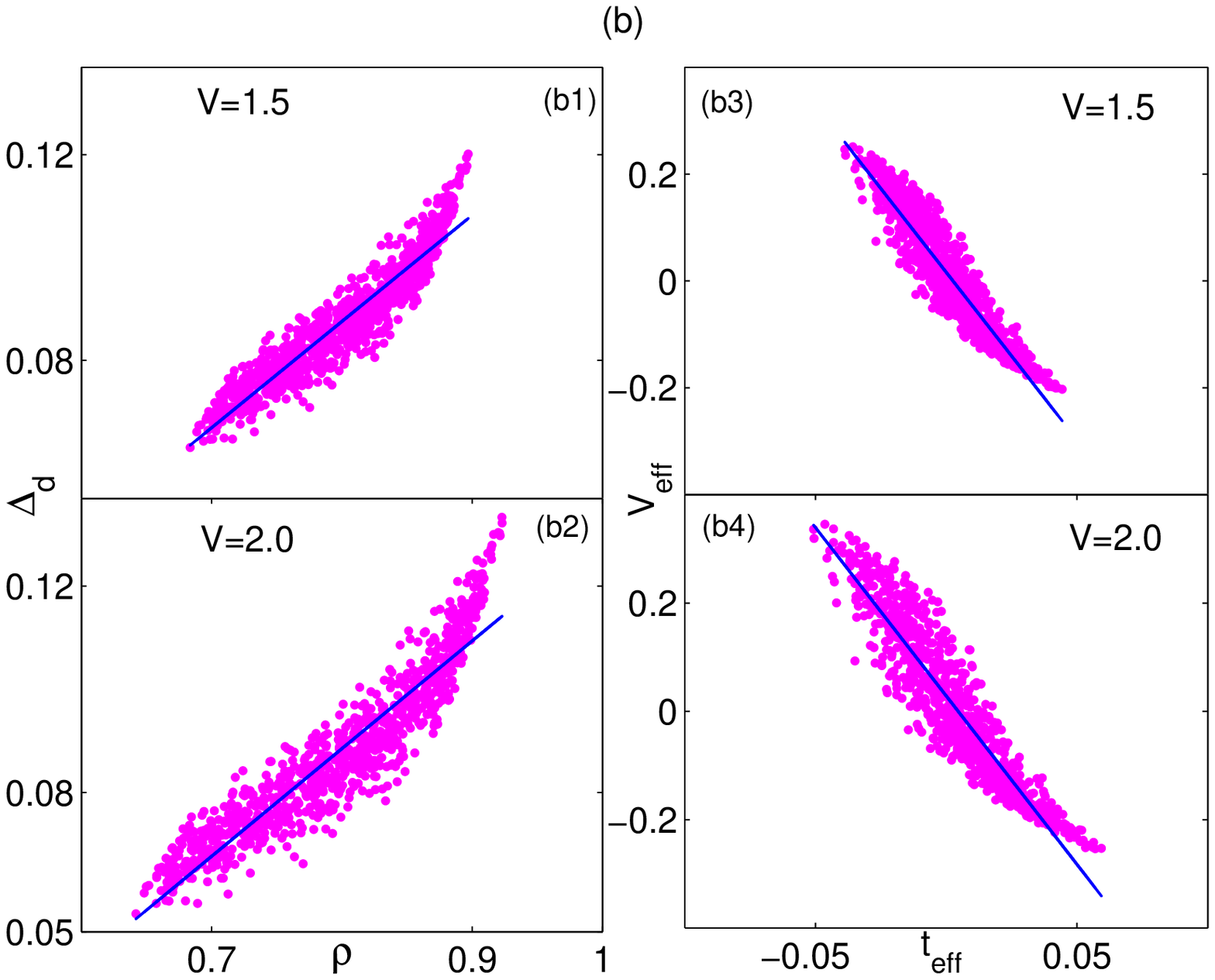}
\end{minipage}
\end{center}
\caption{
(a) The spatial profile of $V_{\rm eff}(i)$ from the GIMT and IMT calculations are shown for $V=1.5$ (a1,a2) and $V=2.0$ (a3,a4). 
The bare $V_i$ is similar to the IMT profiles (not shown separately), as expected from figure~\ref{fig:veffdis}(b). The GIMT results of $V_{\rm eff}(i)$ on the left column of (a) show the evolution of the spatially clustered regions of relatively weaker site-disorder, reflecting the asymmetry of $P_{\rm GIMT}(V_{\rm eff})$ in figure~\ref{fig:veffdis}(a). These are the regions supporting larger $\rho_i$ in figure~\ref{fig:spatialstruc}(b3) and (b5).
(b) The scatter plot of $\Delta_d(i)$ vs $\rho_i$ for the GIMT calculations with $V=1.5$ (b1) and $V=2.0$ (b2) shows the positive spatial correlation between these two quantities (i.e., both being large or small in same locations). The blue lines give a linear fit to the data. (b3) and (b4) show the similar scatter plots of $t_{\rm eff}(i)$ vs $V_{\rm eff}(i)$ from the GIMT scheme indicating the spatial anti-correlation between them.
}
\label{fig:Vscatter}
\end{figure*}

\subsection{Spatially correlated effective disorder}\label{subsec:EffectiveDisorder}

The mean-field Hamiltonian of equation~(\ref{eq:meanfield2}) can be cast in the following effective form:
\begin{eqnarray}
{\cal H}_{\rm MF}  &=& \sum_{i, \delta, \sigma} t_{\rm eff}(i,\delta) ~c^{\dagger}_{i \sigma} c_{i+\delta \sigma} + \sum_{i,\sigma} V_{\rm eff}(i) n_{i\sigma} \nonumber \\
&+& \left\{ {\rm pairing~terms~involving~~} \Delta^{\delta}_i \right \}
\label{eq:tV}
\end{eqnarray}
where the explicit expressions for $t_{\rm eff}$ and $V_{\rm eff}$ can be extracted comparing equation~(\ref{eq:meanfield2}) and (\ref{eq:tV}), both of which become inhomogeneous in the presence of the disorder. The inhomogeneities in $t_{\rm eff}$ from the IMT calculation are very weak though, and that in $V_{\rm eff}$ is largely similar to the bare disorder. Obviously, the physics of the strong correlations within GIMT is also contained in the equation~(\ref{eq:tV}) through local GRF and their derivatives with respect to $\rho_i$. The physical origin for the differences between the IMT and GIMT results can be motivated using the following argument. A strong repulsive interaction smears out any local accumulation of the charge carriers. 
Such a smearing plays an important role in GIMT, but is missing from plain IMT scheme. As a result, the spatial distribution of electrons ought to be different between the two calculations. Such re-organization of the electronic density must modify all other local order parameters in a self-consistent manner and hence should also alter all the physical observables. The question remains: How is this correlation driven physics incorporated in the ${\cal H}_{\rm MF}$? A significant insight is obtained by studying the spatial fluctuations of $t_{\rm eff}$ and $V_{\rm eff}$ individually, which is presented in figure~\ref{fig:tV}. We plot $t_{\rm eff}(i)=\sum_{\delta}t_{\rm eff}(i,\delta)$ and $V_{\rm eff}(i)$ on the lattice both from the IMT and the GIMT calculations. Both the $t_{\rm eff}(i)$ and the $V_{\rm eff}(i)$ show qualitatively similar spatial variations in plain IMT (figure~\ref{fig:tV}(a) and (b)). On the contrary, a strong spatial anti-correlation develops in the fluctuating parts of $t_{\rm eff}$ and $V_{\rm eff}$ within the GIMT scheme as demonstrated in figures~\ref{fig:Vscatter}(b3) and (b4) using scatter plots. Comparison of their spatial profiles in figures~\ref{fig:tV}(c) and (d) shows that $t_{\rm eff}(i)$ is small precisely in those regions where $V_{\rm eff}(i)$ is large and vice versa. Such a competing behavior \cite{Fukushima2008} of the two is a result of the strong repulsive interactions, and we elaborate this point in the following.

According to equation~(\ref{eq:tV}), $t_{\rm eff}(i)$ is the probability of hopping of an electron onto the site $i$ from the neighbors. A site with $V_{\rm eff}(i)>V_{\rm eff}(i+\delta)$ will support $\rho_i<\rho_{i+\delta}$ leading to a larger $g^t_{i,i+\delta}$ (compared to when $V_{\rm eff}$ were same for $i$ and $i+\delta$). This, in turn, enhances $t_{\rm eff}(i)$, increasing the final $\rho_i$ compared to what $V_{\rm eff}(i)$ alone would have predicted in the first place! So high `hills' of disorder profile does not remain as sparsely populated by electrons in GIMT as expected in the IMT calculation. Exactly opposite mechanism smears out the charge accumulation in GIMT from sites with deep `valleys' of disorder potential.
  
On the other hand, $t_{\rm eff}(i)$ is largely independent of $\rho_i$ within plain IMT framework, though it develops a very weak density-dependence through the Fock-shifts. In fact, $t_{\rm eff}(i)$ becomes weaker in regions with high-hills and deep-valleys of disorder profile within plain IMT. Such sites tend to get decoupled from hopping in the absence of the `feedback-loop' in terms of $g^t_{i,i+\delta}$, quite in contrast to the GIMT scenario discussed above.
This picture is substantiated through our results in figures~\ref{fig:tV}(e) and (f), where we demonstrate the competing effects of $t_{\rm eff}(i)$ and $V_{\rm eff}(i)$ on $\rho_i$, if the two were to work individually. It is thus no surprise that the GIMT local density (Left column, figure~\ref{fig:spatialstruc}(b), incorporating both $t_{\rm eff}$ and $V_{\rm eff}$) becomes less inhomogeneous than the IMT finding.

\begin{figure*}
\begin{center}
\begin{minipage}{.9\columnwidth}
\includegraphics[width=1.0\columnwidth,clip=]{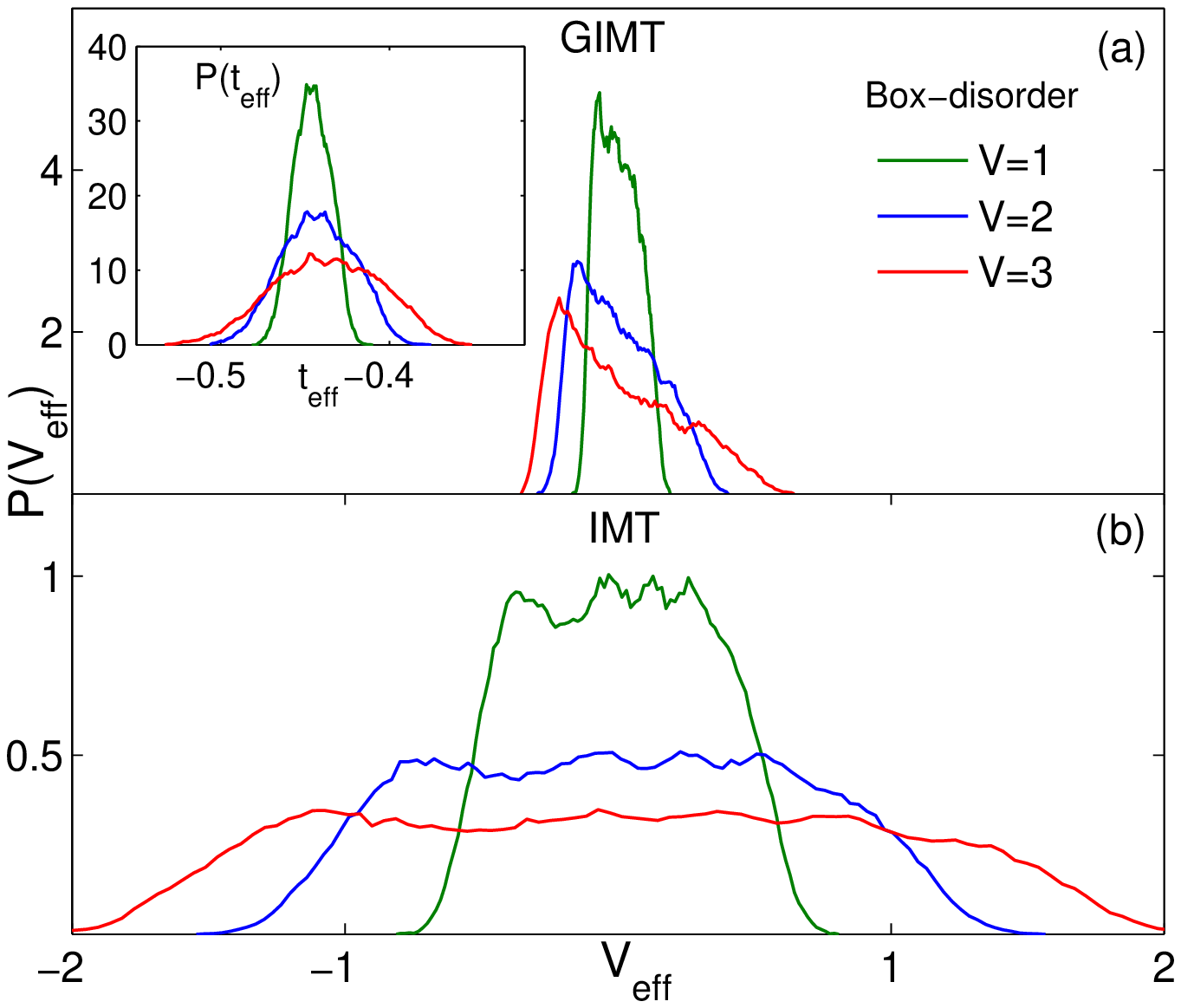}
\end{minipage}
\begin{minipage}{.9\columnwidth}
\includegraphics[width=1.0\columnwidth,clip=]{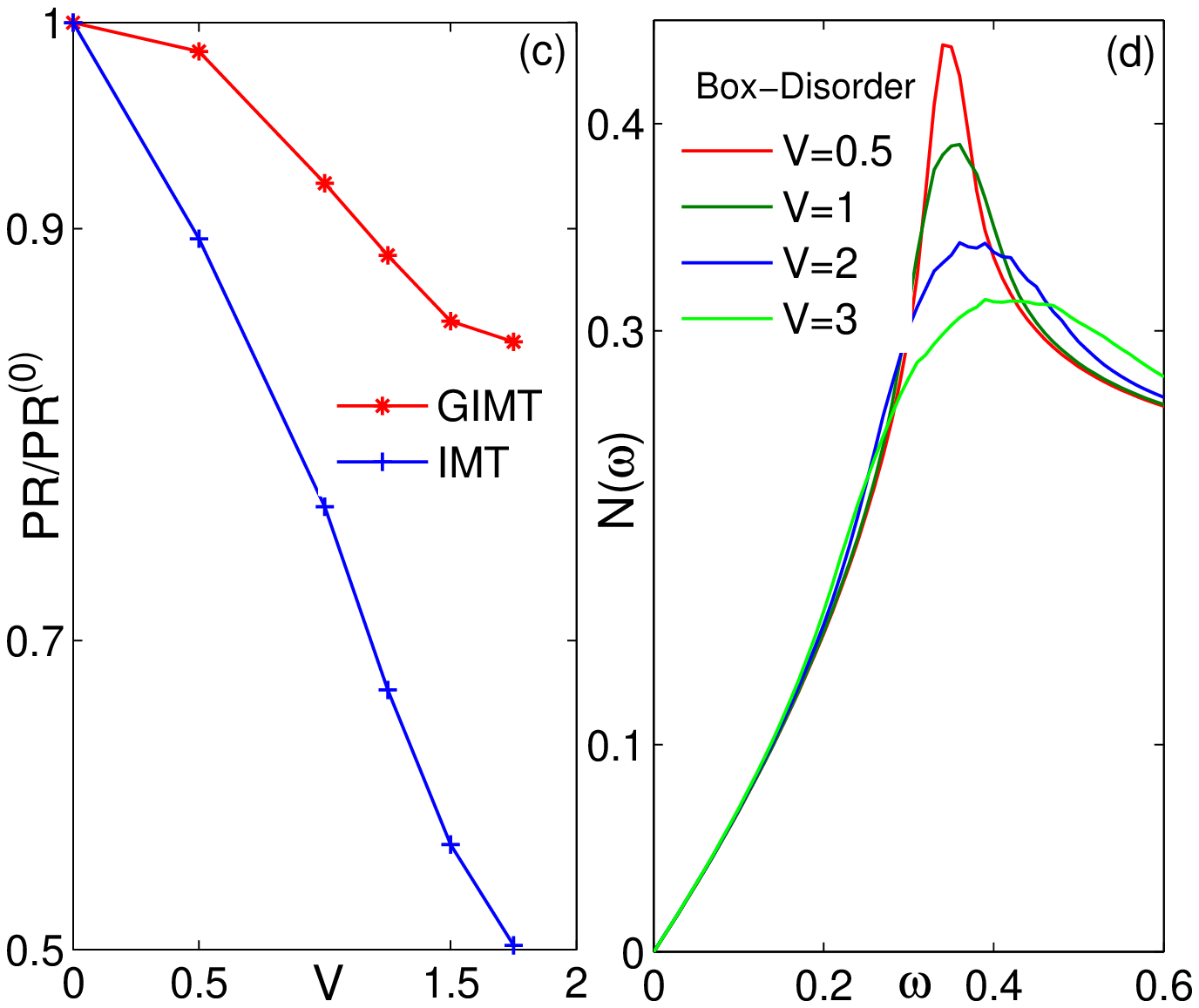}
\end{minipage}
\end{center}
\caption{The distribution $P(V_{\rm eff})$ of the effective disorder $V_{\rm eff}(i)$, are presented in (a) from the GIMT and in (b) from IMT  calculations. The homogeneous components of $V_{\rm eff}$ (from the Hartree-shift at $V=0$) are $1.52$ and $-0.87$ respectively in the IMT and GIMT results, are subtracted for clarity. (a) The narrow $P_{\rm GIMT}(V_{\rm eff})$ and its intriguing asymmetry implies that there are more sites with `valleys' than `hills' in $V_{\rm eff}(i)$, whereas the two are balanced in {\it bare} $V_i$ by construction. This asymmetry directly connects to the opposite asymmetry in $P(\rho)$ of figure~\ref{fig:Pden}(a). The inset with $P_{\rm GIMT}(t_{\rm eff})$ shows significant fluctuations in $t_{\rm eff}$ relative to its homogeneous value. (b) $P_{\rm IMT}(V_{\rm eff})$ highlights its contrast with $P_{\rm GIMT}(V_{\rm eff})$ on the following grounds: (i) The $V_{\rm eff}$ from the IMT {\it increases} compared to the bare one ($|V_i| \leq  V/2$). (ii) The symmetry of the bare $P(V_i)$ between positive and negative values is maintained in $P_{\rm IMT}(V_{\rm eff})$, though the edges of the box-disorder smear out.
(c) The evolution of the participation ratio (PR)
is compared between the normal states (NS) of the GIMT and IMT findings. The NS from the GIMT remains largely extended up to $V=1.75$, and begins to show the signature of a saturation. In contrast, the PR from the IMT decreases steadily yielding localized NS at large $V$. While the ${\rm PR}_{\rm IMT}$ continues to fall up to $V \sim 3$ (not shown here), we encountered severe convergence issues with ${\rm NS}_{\rm GIMT}$ at large $V$. In fact, it is harder to converge to ${\rm NS}_{\rm GIMT}$ than to the full superconducting GS.
(d) The average $N(\omega)$ obtained from equation~(\ref{eq:larkindos}), with $P(\Delta_d)$ taken directly from the GIMT results of figure~\ref{fig:Pden}(c).
The strong resemblance of the resulting $N(\omega)$, particularly for small $|\omega|$, with that in figure~\ref{fig:dos}(a) demonstrates the irrelevance of the spatial inhomogeneity.
}
\label{fig:veffdis}
\end{figure*}

Finally, we note that one of the important effects of the strong correlation lies in the significant reduction \cite{Fukushima09} of the magnitude of $V_{\rm eff}(i)$ from the bare $V_i$. We demonstrate this by presenting the distribution of $P(V_{\rm eff})$ in figures~\ref{fig:veffdis}(a) and \ref{fig:veffdis}(b) both from the GIMT and plain IMT calculations. The homogeneous component of $V_{\rm eff}$ (the contribution at $V=0$) from the Hartree-shift is already subtracted in this analysis. 
We see that while in the IMT calculations the range of $V_{\rm eff}(i)$ actually increases a bit from the bare value of $\pm V/2$, strong electronic correlations indeed cause its drastic reduction as seen from the resulting $P_{\rm {GIMT}}(V_{\rm eff})$. Interestingly, $P_{\rm {GIMT}}(V_{\rm eff})$ becomes more asymmetric with increasing $V$ with larger weight accumulating for negative $V_{\rm eff}(i)$. In contrast, the $P_{\rm {IMT}}(V_{\rm eff})$ maintains its symmetry around zero just as in the bare $P(V_i)$ by construction. Nevertheless, we found that the asymmetric $P_{\rm GIMT}(V_{\rm eff})$ still produces $\langle V_{\rm eff} \rangle=0$ for all $V$. In fact, it is this asymmetry that translates into an oppositely asymmetric $P_{\rm GIMT}(\rho_i)$ in figure~\ref{fig:Pden}(a). We comprehend this intriguing feature in the following way: For a non-interacting problem at large $V$, the local density ranges from zero to double-occupancy on the highest hill and the deepest valley, respectively. The GA, however, restricts $\rho_i \leq 1$, breaking the symmetry of the non-interacting problem. A self-consistent interplay of the strong correlations and the disorder now ensures that the effective disorder has more `attractive' points such that the maximal $\rho_i$ is limited by half-filling everywhere, still maintaining the global density at the desired value ($\rho = 0.8$, in our case). An evolution  of spatial correlation in $V_{\rm eff}(i)$ with disorder, producing larger patches of attractive points is further illustrated using colour-density plots in figure~\ref{fig:Vscatter}(a), indicating a growing spatial correlation of such attractive points. Let us emphasize that we verified the crucial physical insights, emerging from the study of the spatial distribution of the local observables, survive in all the realizations of disorder we considered. 

\subsection{Underlying normal state: Anderson's theorem for cuprates?}\label{subsec:Normal}

The intriguing evolution of the spatially correlated effective disorder in the GIMT calculations immediately tells us that they must also modify the properties of the underlying normal state (NS). Here, the NS is defined as the solution of the same $\cal{H}_{\rm MF}$ of equation~(\ref{eq:tV}), but suppressing all the Cooper-channel order parameters $\Delta^{\delta}_i$, for all $i$ and $\delta$. We caution the readers that our usage of the term `normal state' might lead to a confusion, because it is frequently used to identify the $T>T_c$ phase of a HTSC, where the d-wave order is already destroyed by the thermal fluctuations. Our NS, on the other hand, is a true GS at $T=0$, had there been no pairing. As the locally renormalized GRF leads to spatially correlated disorder, the ${\rm NS}_{\rm GIMT}$ must differ from the ${\rm NS}_{\rm IMT}$, where the later is essentially the solution of the Anderson model \cite{Anderson58} of non-interacting electrons in disordered background. The ${\rm NS}_{\rm GIMT}$ plays a crucial role for comprehending our GIMT results hinting towards a physical picture for the dirty cuprates, as we discuss below.

The robustness of the observables to the impurities, as demonstrated in section~\ref{subsec:PhysObs}, is actually reminiscent of Anderson's theorem \cite{Anderson59}. While the theorem does not apply to a weak-coupling d-wave superconductor \cite{Abrikosov61}, our results open up such a possibility for strongly correlated cuprates. In order to support this scenario, we first argue that our GIMT results are consistent with Anderson's mechanism provided we extend the scope of pairing between all the `correlated' single-particle normal states (${\rm NS}_{\rm GIMT}$). To establish this, we solve for ${\cal H}^{\rm Normal}_{\rm MF}$ (equation~(\ref{eq:tV}) with all $\Delta^{\delta}_i$ terms suppressed) to obtain the ${\rm NS}_{\rm GIMT}$ parameters $t_{\rm eff}(i)$ and $V_{\rm eff}(i)$ in a self-consistent manner in the first step. These parameters carry information on the local orders in and Hartree- and Fock-channels. In the next step, we set up a BdG-type matrix $\tilde{{\cal H}}_{\rm BdG}$ in terms of these ${\rm NS}_{\rm GIMT}$ parameters in the diagonal blocks, and with $\Delta^{\delta}_i$ in the off-diagonal blocks, where $\Delta^{\delta}_i$ are obtained from the corresponding GIMT solution of the full ${\cal H}_{\rm MF}$, including the dSC order. This $\tilde{{\cal H}}_{\rm BdG}$ is then diagonalized only {\it once} in the final step, and we recalculate the new (non self-consistent) local pairing amplitudes $\tilde{\Delta}^i_{\delta}$ from this sole diagonalization. We find an impressive agreement between $\Delta^i_{\delta}$ and $\tilde{\Delta}^i_{\delta}$ for all $i$ and $\delta$. Such a non self-consistent calculation is equivalent to an Anderson-type idea (though the pairing in this case is not limited to the time-reversed states alone).
This interpretation works because the effective attraction does not seem to renormalize the spatially fluctuating NS orders in the diagonal blocks of $\tilde{{\cal H}}_{\rm BdG}$. If instead, we had any significant mismatch between $\Delta^i_{\delta}$ and $\tilde{\Delta}^i_{\delta}$, a feedback loop would have been required for self-consistency which could, in general, update NS order parameters as well, implying that the pairing modifies single-particle states. Such a conclusion goes against Anderson's philosophy, according to which pairing from effective attraction is independent of disorder. This is because, the impurities are already consumed in producing the corresponding ``exact eigenstates".  
Note that in the case of the sSC, the normal states are essentially the exact eigenstates of the effective single-particle Hamiltonian, as there are no strong correlations.

\begin{figure*}[t]
\begin{center}
\begin{minipage}{.91\columnwidth}
\includegraphics[width=1.0\columnwidth,clip=]{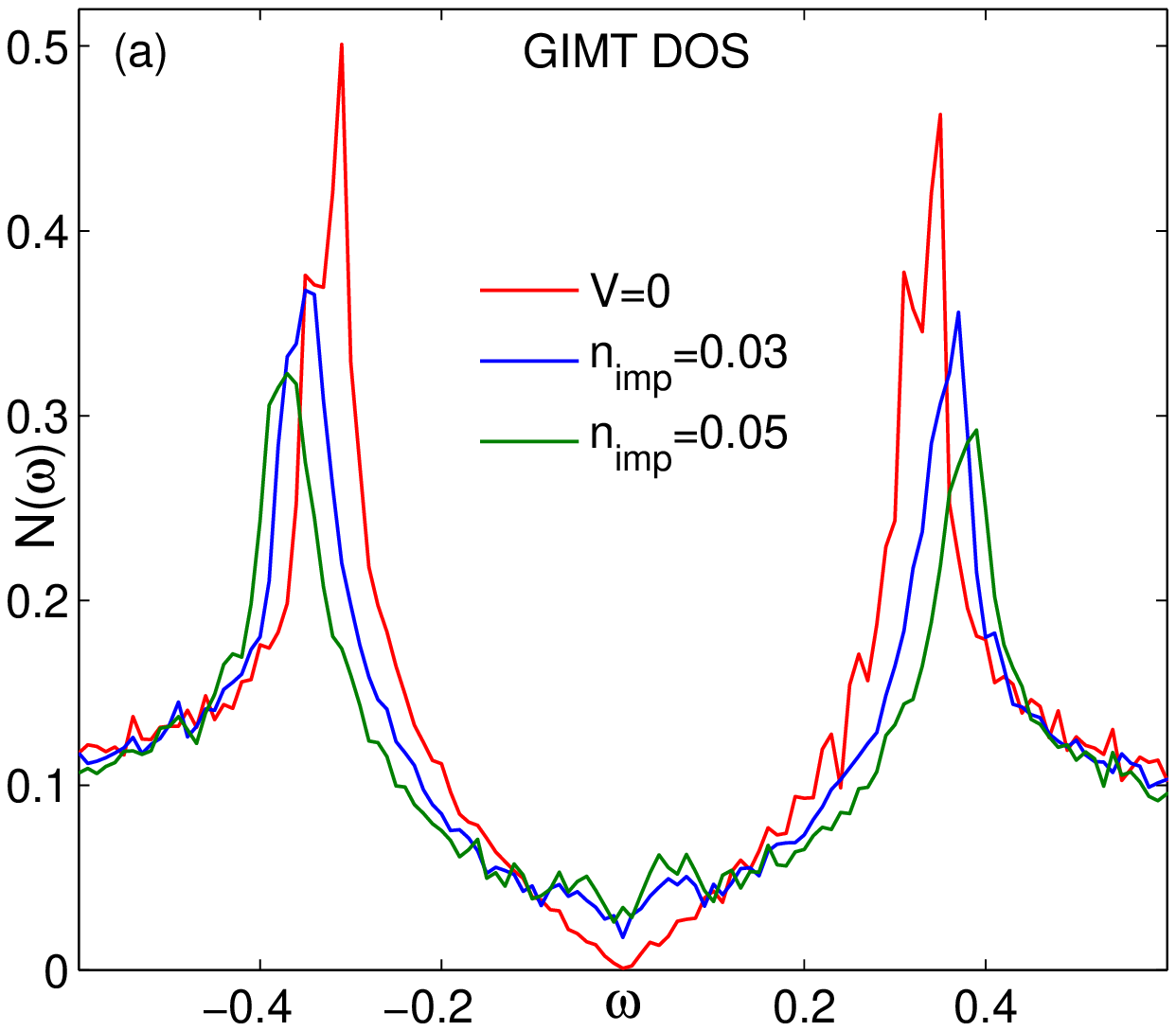}
\end{minipage}
\begin{minipage}{.87\columnwidth}
\includegraphics[width=1.0\columnwidth,clip=]{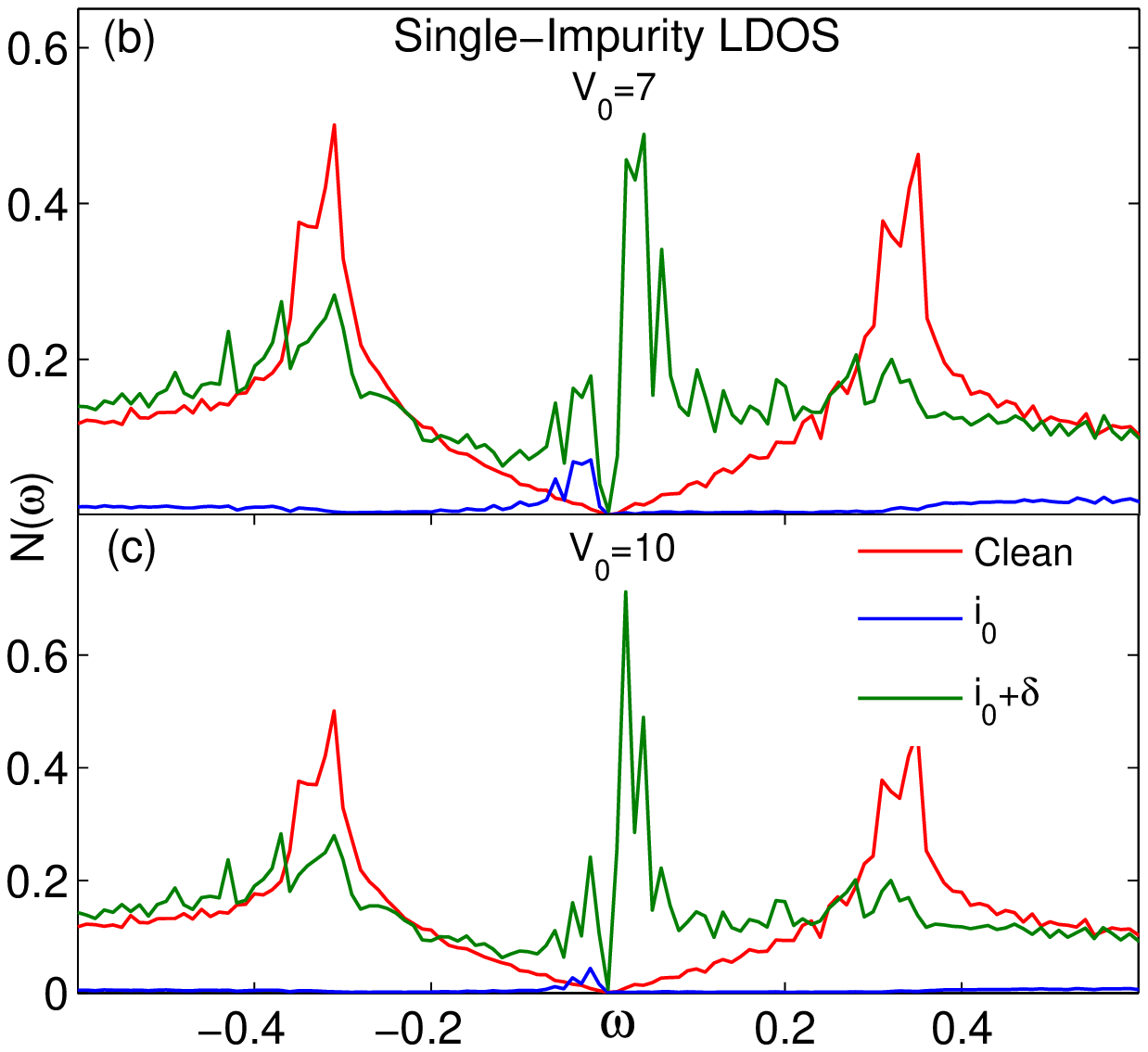}
\end{minipage}
\end{center}
\caption{
(a) The $N(\omega)$ in the limit of strong $V_0$ (conc-disorder) begins to show the signatures of a gap-filling. The site-averaged $N(\omega)$ for $V_0=5$ where some accumulation of mid-gap states are evident for $n_{\rm imp}=0.03, 0.05$. Results from a single impurity calculation with (b) $V_0=7$ and (c) $V_0=10$ showing LDOS on the impurity site $i_0$ and also on $i_0+\delta$. Such a gap-filling heals quickly over a short distance, and LDOS follows the profile at $V=0$ on the sites beyond the next-nearest neighbor from $i_0$.
$N(i_0+\delta,\omega \simeq 0)$ was found to increase with $V_0$.
}
\label{fig:strong}
\end{figure*}

It is not only the pairing between the normal states which is responsible for the insensitivity of superconductivity to the impurities. The extent of localization of these states is also crucial for the inferences made by Anderson's theorem as we discuss in the following. While such a pairing theory for a sSC can be formulated in principle for any strength of disorder tailoring the disorder-induced inhomogeneities in it, the robustness of s-wave superconductivity breaks down at large disorder ($V \geq t$) due to strong localization \cite{Ghosal01}. It is interesting to note that the insensitivity of observables for dirty cuprates actually extends even beyond $V > 2.5$ (See figures~\ref{fig:odlro} and \ref{fig:dos}), a strength by which traditional Anderson's theorem (for sSC) weakens drastically. In order to settle this apparent conflict, we study the differences between the NS from the GIMT and IMT schemes, focusing on their localization properties.
We quantify the extent of localization by calculating the participation ratio (PR) for the effective one-particle NS near $\mu$, defined as:
\begin{equation}
(PR)_n=\sum_{i} \left| \psi_n(i) \right|^4
\label{eq:pr}
\end{equation}
where $\psi_n(i)$ is the normalized eigenfunctions of the NS for the eigenvalue $\varepsilon_n \approx \mu$. For a better statistics, we actually average PR over all those $n$'s such that $|\varepsilon_n - \mu| \leq 0.03$. The PR measures the fraction of sites of the system contributing to the $n$th wave function, and is a good estimate of the localization area in two-dimensions. The behavior of PR with the disorder strength $V$ in figure~\ref{fig:veffdis}(c) shows that the one-particle ${\rm NS}_{\rm GIMT}$ (with $\varepsilon_n \approx \mu$) remains fairly extended up to $V=1.75$ on a scale of the system size. In contrast, the ${\rm NS}_{\rm IMT}$, pairing of which leads to the Anderson's theorem for sSC, gets localized relatively fast! The normal state DOS, $N_{\rm NS}(\omega \approx \mu)$, that incorporates the GA was found to be insensitive to $V$ (though shows a strong sample to sample fluctuations) \cite{Ma85}, consistent with the extended nature of the ${\rm NS}_{\rm GIMT}$. The robustness of observables for dirty superconductors follows as a consequence of traditional pairing philosophy of Anderson, as long as the exact eigenstates (or NS in our terminology) remains fairly extended in the scale of the system size. A similar picture emerges for the dirty cuprates  from the above discussions: (a) The GIMT findings can be thought of as a result of pairing between correlated normal states. (b) Such normal states remain delocalized up to a large disorder.

An important consequence of the conventional Anderson's theorem \cite{Anderson59} is the irrelevance of any detailed spatial information of the system. Such irrelevance for our case is seen by calculating $\tilde{N}(\omega)$ using the following prescription \cite{Larkin72}:
\begin{equation}
\tilde{N}(\omega)=\int d\Delta_d P(\Delta_d) N_{\rm BCS}(\omega,\Delta_d)
\label{eq:larkindos}
\end{equation}
where, $N_{\rm BCS}$ is the density of states of a homogeneous d-wave BCS superconductor and is given by \cite{Maki94}
\begin{eqnarray}
N_{\rm BCS}(\omega,\Delta_d)&=&\frac{2}{\pi}\frac{|\omega|}{\Delta_d}{\cal K}\Big(\frac{|\omega|}{\Delta_d}\Big)~~\rm {for}~~\frac{|\omega|}{\Delta_d}\leq 1, \nonumber \\
&=&\frac{2}{\pi}{\cal K}\Big(\frac{\Delta_d}{|\omega|}\Big)~~\rm {for}~~\frac{|\omega|}{\Delta_d}> 1.
\label{eq:homdos}
\end{eqnarray}
Here, ${\cal K}(x)$ is the complete elliptic integral and $\Delta_d$ being the d-wave gap at $T=0$. Such a definition keeps track of the fluctuations in the pairing amplitude only on an average and no information about the spatial structures of inhomogeneities is accounted for. The resulting $\tilde{N}(\omega)$, using $P(\Delta_d)$ from our GIMT calculation, is presented in figure~\ref{fig:veffdis}(d). We show $\tilde{N}(\omega)$ only for the positive $\omega$ because it is manifestly symmetric about $\omega=0$.  The similarity of $\tilde{N}(\omega)$ with $N_{\rm {GIMT}}(\omega)$ of figure~\ref{fig:dos}(a) illustrates the redundancy of any spatial correlation of inhomogeneities in $N_{\rm {GIMT}}(\omega)$.

Before closing the current discussion, we must mention that the above arguments supporting the robustness of cuprate superconductors as being similar to the conventional Anderson's theorem, are all {\it indirect} ones, though they add up to a significant claim! It will be interesting to substantiate these claims also by direct demonstration. Such a formalism is beyond the scope of this study and will be addressed elsewhere.

It is important to address the fate of the cuprates as the disorder strength is increased beyond what is reported here. 
It is easy to see that the GIMT scheme, which removes {\it all} double-occupancy by enforcing $\rho_i \leq 1$ (for all $i$), would not work on the sites with $V_i\geq U$.
However this is not a deficit of the GA. It would only require a modification of allowing occasional $\rho_i > 1$ based on relative strength of $U$ and $V_i$. Such calculations, however, are beyond the scope of our study and will be addressed elsewhere. Nevertheless, we found that the GIMT self-consistency cracks up technically even for $V > 3.0$ (box-disorder). In fact, making it work for large $V$ requires significant improvements in the iterative self-consistency in terms of applying accelerated convergence methods. Our usage of Broyden \cite{Johnson88} and modified Broyden \cite{Knapp05} methods was crucial in pushing the self-consistency to larger disorder strengths.

It turns out that conc-disorder helps to push the self-consistent calculations towards a higher $V_0$, because we could tune $n_{\rm imp}$ independently to keep iterative self-consistency under control. The average DOS calculated with $V_0=5$ and $n_{\rm imp}=0.03$, $0.05$ are presented in figure~\ref{fig:strong}(a) which shows the first evidence of a gap-filling within the GIMT scheme \cite{Andersen08}. Figures~\ref{fig:strong}(b) and \ref{fig:strong}(c) present the LDOS from the GIMT calculation around a single impurity, where the strength of the sole impurity is $V_0=7$ and $V_0=10$, an energy scale larger than $t$, $t'$, $J$ and $J'$. 
Results for both $V_0=7$ and $V_0=10$ show that growing $V_0$ produces larger $N({\mathbf r},\omega \simeq 0)$, where ${\mathbf r}$ represents the nearest neighbors of the impurity site. A simple picture emerges from repeating the calculations for several combinations of $V_0$ and $n_{\rm imp}$: The gap filling begins when $V_0$ grows beyond the bandwidth of the homogeneous, Gutzwiller-renormalized normal state, i.e., $V_0 > 8g^tt$, where $g^t=(1-\rho)/(1-\rho/2)$. This observation motivates the formalism of an alternate IMT scheme elaborated further in the supplementary material.

\section{Conclusion}\label{sec:conclu}

In conclusion, we have studied the qualitative and quantitative effects of the disorder-induced inhomogeneities on a strongly correlated dSC. We established that such fluctuations in the spatial orders are fundamentally different from when strong correlations are neglected. Such differences in GIMT do not allow formation of superconducting-`islands', though produces nanoscale inhomogeneities. The surprising insensitivity of $\Delta_{\rm OP}$, $D_s$ and $N(\omega)$ to impurities is reminiscent of Anderson's theorem, allowing a simple understanding of the complex physics of dirty cuprates.
Even though the spatial inhomogeneity in the order parameters are somewhat weaker from plain IMT, the outcome of our GIMT results remain substantially different from the prediction of the Abrikosov-Gorkov theory or self-consistent T-matrix approximation.
The additional effects of strong correlations in the GIMT scheme make the underlying normal states substantially different from the conventional metallic ones.
The GIMT normal states differ from the localized phase of the Anderson model because it generates spatially correlated effective disorder. The superconducting GS within the GIMT scheme is interpreted as a result of pairing between these correlated normal states. While a large part of the differences between the GIMT and IMT can be attributed to the changes in the effective disorder, they cannot complete the story just by themselves. A complex and correlated interplay of the effective disorder and pairing modifies the scenario in a self-consistent manner.
Our results from the GIMT method help comprehending much of the experimental trends. Extension of the GIMT scheme to variationally include inhomogeneous double-occupancy (consistent with local disorder) is an important future direction. It will be interesting to consider the effect of quantum phase fluctuations. While they play an important role on a dSC \cite{Arun01}, our GIMT results of the lesser fluctuation in the pairing amplitude call for revisiting the effects of the quantum phase fluctuations riding on top of the GIMT state in a dSC. We speculate that our results will motivate further research to address surprises from the dirty superconductors.    

\section*{Acknowledgements}
We thank A. Garg for valuable conversations. DC acknowledges his fellowship from CSIR (India).

\appendix
\section{Detailed expressions for GRFs used in section~\ref{subsec:GA}}\label{sec:appendix}
In this appendix, we provide the formulae used in section~\ref{subsec:GA}. Carrying out the analysis outlined by equations~(\ref{eq:energy}) and (\ref{eq:meanfield1}), we find the form of $G_1$ to $G_5$ in equation~(\ref{eq:meanfield2}) as,
\begin{eqnarray}
&&G_1^{i, \delta}=-\frac{J}{4} \left\{ \Bigg( \frac{3g_{i,i+\delta}^{xy}+1}{4} \Bigg)-\Bigg( \frac{g_{i,i+\delta,i}^{31}+g_{i,i+\delta,i}^{32}}{2} \Bigg) \right\}, \nonumber \\
&&G_2^{i, \delta, \delta'}=-\frac{J}{4} \Bigg( \frac{g_{i+\delta,i,i+\delta'}^{31}+g_{i+\delta,i,i+\delta'}^{32}}{2} \Bigg),  \nonumber \\
&&G_3^{i,\delta,\tilde{\delta}}=\frac{tt'}{U} ( g_{i+\delta,i,i+\tilde{\delta}}^{31}+g_{i+\delta,i,i+\tilde{\delta}}^{32} ), \nonumber \\
&&G_4^{i,\delta,\delta'}=-\frac{J}{8} g_{i+\delta,i,i+\delta'}^{31} \rho_i,~G_5^{i,\delta,\tilde{\delta}}=\frac{tt'}{U} \rho_i
\end{eqnarray}
with $G_n^{i,\tilde{\delta},(\tilde{\delta}')}=G_n^{i,\delta,(\delta')}(J \rightarrow J',\delta \rightarrow \tilde{\delta})$ for $n=1,2,4$ and $(\tilde{\delta}')$ is included in the superscript, wherever applicable. The Fock-shifts in equation~(\ref{eq:meanfield2}) are given by,
\begin{eqnarray}
W_{i \delta}^{FS}&=&\frac{J}{2} \left\{ \Bigg( \frac{3g_{i,i+\delta}^{xy}}{2} -\frac{g_{i,i+\delta,i}^{32}}{2}-\frac{g_{i+\delta,i,i+\delta}^{32}}{2} -\frac{1}{2} \Bigg)  \tau_{i}^{\delta} \right\}, \nonumber \\
W_{i \tilde{\delta}}^{FS}&=&W_{i \delta}^{FS}(J \rightarrow J',\delta \rightarrow \tilde{\delta})
\end{eqnarray}
where $\tau_{i}^{\delta}$ ($\equiv \tau_{ij}$) lives on the bonds, here $j=i+\delta$. The Hartree-shift in equation (\ref{eq:meanfield2}) gets its major contribution from the derivatives of GRF with respect to local density:
\begin{eqnarray}
&&\mu_i^{HS}=-4t\sum_{\delta,\sigma} \left\{ \frac{\partial{g_{i,i+\delta}^{t}}}{\partial{n_i}} \tau_i^{\delta} \right\}+4t'\sum_{\tilde{\delta},\sigma} \left\{ \frac{\partial{g_{i,i+\tilde{\delta}}^{t}}}{\partial{n_i}} \tau_i^{\tilde{\delta}} \right\} \nonumber \\ 
&&-\frac{3J}{2}\sum_{\delta,\sigma} \frac{\partial{g_{i,i+\delta}^{xy}}}{\partial{n_i}}\Big( \frac{1}{4} {\Delta_i^{\delta}}^2+{\tau_i^{\delta}}^2 \Big) \nonumber \\
&&-\frac{3J'}{2}\sum_{\tilde{\delta},\sigma} \frac{\partial{g_{i,i+\tilde{\delta}}^{xy}}}{\partial{n_i}}\Big( \frac{1}{4} {\Delta_i^{\tilde{\delta}}}^2+{\tau_i^{\tilde{\delta}}}^2 \Big) \nonumber \\ 
&& -J  \sum_{\delta,\delta',\sigma \atop \delta \neq -\delta'} \left\{ \Bigg( \frac{\partial{g_{i,i+\delta,i+\delta+\delta'}^{31}}}{\partial{n_i}} + \frac{1}{4} \frac{\partial{g_{i,i+\delta,i+\delta+\delta'}^{32}}}{\partial{n_i}} \Bigg) \Delta_i^{\delta}\Delta_{i+\delta}^{\delta'} \right. \nonumber \\
&&\left. +\frac{1}{2} \frac{\partial{g_{i,i+\delta,i+\delta+\delta'}^{31}}}{\partial{n_i}} \tau_i^{\delta+\delta'} \rho_{i+\delta} +\frac{\partial{g_{i,i+\delta,i+\delta+\delta'}^{32}}}{\partial{n_i}} \tau_i^{\delta} \tau_{i+\delta}^{\delta'} \right. \nonumber \\
&& \left. + \sum_{\delta,\delta',\sigma \atop \delta \neq \delta'} \frac{1}{2} \frac{\partial{g_{i+\delta,i,i+\delta'}^{32}}}{\partial{n_i}} \Big( \frac{1}{4} \Delta_i^{\delta} \Delta_i^{\delta'} + \tau_i^{\delta} \tau_i^{\delta'} \Big) \right\} \nonumber \\
&& -J' \sum_{\tilde{\delta},\tilde{\delta}',\sigma \atop \tilde{\delta} \neq -\tilde{\delta}'} \left\{  \Bigg( \frac{\partial{g_{i,i+\tilde{\delta},i+\tilde{\delta}+\tilde{\delta}'}^{31}}}{\partial{n_i}} + \frac{1}{4} \frac{\partial{g_{i,i+\tilde{\delta},i+\tilde{\delta}+\tilde{\delta}'}^{32}}}{\partial{n_i}} \Bigg) \Delta_i^{\tilde{\delta}}\Delta_{i+\tilde{\delta}}^{\tilde{\delta}'} \right. \nonumber \\
&&\left. +\frac{1}{2} \frac{\partial{g_{i,i+\tilde{\delta},i+\tilde{\delta}+\tilde{\delta}'}^{31}}}{\partial{n_i}} \tau_i^{\tilde{\delta}+\tilde{\delta}'} \rho_{i+\tilde{\delta}} +\frac{\partial{g_{i,i+\tilde{\delta},i+\tilde{\delta}+\tilde{\delta}'}^{32}}}{\partial{n_i}} \tau_i^{\tilde{\delta}} \tau_{i+\tilde{\delta}}^{\tilde{\delta}'} \right. \nonumber \\
&& \left. + \sum_{\tilde{\delta},\tilde{\delta}',\sigma \atop \tilde{\delta} \neq \tilde{\delta}'} \frac{1}{2} \frac{\partial{g_{i+\tilde{\delta},i,i+\tilde{\delta}'}^{32}}}{\partial{n_i}} \Big( \frac{1}{4} \Delta_i^{\tilde{\delta}} \Delta_i^{\tilde{\delta}'} + \tau_i^{\tilde{\delta}} \tau_i^{\tilde{\delta}'} \Big) \right\} \nonumber \\
&&+ \frac{4tt'}{U} \sum_{i,\delta,\tilde{\delta},\sigma}\left\{ \frac{\partial{g_{i+\delta,i,i+\tilde{\delta}}^{32}}}{\partial{n_i}} \Big( \frac{1}{2} \Delta_i^{\delta} \Delta_i^{\tilde{\delta}} + 2 \tau_i^{\delta} \tau_i^{\tilde{\delta}} \Big) \right. \nonumber \\
&& \left. +\Big( \frac{\partial{g_{i,i+\delta,i+\delta+\tilde{\delta}}^{31}}}{\partial{n_i}}+\frac{\partial{g_{i,i+\delta,i+\delta+\tilde{\delta}}^{32}}}{\partial{n_i}} \Big) \frac{1}{2} \Delta_i^{\delta} \Delta_{i+\delta}^{\tilde{\delta}} \right. \nonumber \\
&& \left. +\frac{\partial{g_{i,i+\delta,i+\delta+\tilde{\delta}}^{31}}}{\partial{n_i}} \tau_i^{\delta+\tilde{\delta}}\rho_{i+\delta} +\frac{\partial{g_{i,i+\delta,i+\delta+\tilde{\delta}}^{32}}}{\partial{n_i}} 2 \tau_i^{\delta}\tau_{i+\delta}^{\tilde{\delta}} \right. \nonumber \\
&& \left. + \Big( \frac{\partial{g_{i+\delta+\tilde{\delta},i+\delta,i}^{31}}}{\partial{n_i}}+\frac{\partial{g_{i+\delta+\tilde{\delta},i+\delta,i}^{32}}}{\partial{n_i}} \Big) \frac{1}{2} \Delta_{i+\tilde{\delta}}^{\delta} \Delta_{i}^{\tilde{\delta}} \right. \nonumber \\
&& \left. \frac{\partial{g_{i+\delta+\tilde{\delta},i+\delta,i}^{31}}}{\partial{n_i}} \tau_{i}^{\delta+\tilde{\delta}} \rho_{i+\tilde{\delta}} +\frac{\partial{g_{i+\delta+\tilde{\delta},i+\delta,i}^{32}}}{\partial{n_i}} 2 \tau_{i+\tilde{\delta}}^{\delta} \tau_i^{\tilde{\delta}} \right\} \nonumber \\
&&-\frac{J}{4} \sum_{i,\delta,\sigma} \Big( 1-\frac{g_{i,i+\delta,i}^{31}}{2}-\frac{g_{i+\delta,i,i+\delta}^{31}}{2} \Big) \rho_{i+\delta} \nonumber \\
&&-\frac{J'}{4} \sum_{i,\tilde{\delta},\sigma} \Big( 1-\frac{g_{i,i+\tilde{\delta},i}^{31}}{2}-\frac{g_{i+\tilde{\delta},i,i+\tilde{\delta}}^{31}}{2} \Big) \rho_{i+\tilde{\delta}} \nonumber \\
&&-\frac{J}{4} \sum_{i,\delta,\delta' \atop \sigma}g_{i+\delta,i,i+\delta'}^{31} \tau_{i+\delta}^{\delta'-\delta}-\frac{J'}{4} \sum_{i,\tilde{\delta},\delta' \atop \sigma}g_{i+\tilde{\delta},i,i+\tilde{\delta}'}^{31} \tau_{i+\tilde{\delta}}^{\tilde{\delta}'-\tilde{\delta}}
\end{eqnarray}
Evaluation of equation (A.3) with the cuprate parameters ensures that the major contributions to $\mu_i^{\rm HS}$ come from the derivatives of the GRFs.

\section{Alternative scheme to calculate ODLRO}\label{sec:appendix2a}

Calculation of $\Delta_{\rm OP}$ by evaluating the four-fermionic correlator in equation~(\ref{eq:odlro}) (see section~\ref{subsec:ODLRO}) is numerically expensive. An alternate route for obtaining $\Delta_{\rm OP}$ is to use: $\tilde{\Delta}_{\rm OP}=\langle\Delta_d(i) \rangle \equiv \int d\mathbf{r} \Delta_d(\mathbf{r})P(\Delta)$. An equality of $\Delta_{\rm OP} \approx \tilde{\Delta}_{\rm OP}$ is easily rationalized at $V=0$ dictated by the homogeneity, and also for large $V$ when the spatial correlation of $F_{\delta, \delta'}(i-j)$ at site $i$ and $j$ is quickly lost for $|i-j| > \xi_{\rm loc}$ ($\xi_{\rm loc}$ being the localization length of the corresponding non-interacting system). The validity of the above equality at the two extreme limits of $V$ prompts us to use the conjecture $\Delta_{\rm OP} \approx \langle\Delta_d(i) \rangle$ for all $V$, which has also been verified in other studies \cite{Ghosal01,Jiang13}. We have found this to remain valid, modulo the fact that appropriate GRFs are incorporated, e.g., $\Delta_{\rm OP} \approx \tilde{\Delta}_{\rm OP} = N^{-1}\sum_{i,\delta} g^{t}_{i,i+\delta} (-1)^{\delta_y} \Delta_{i}^{\delta}$.

\section{Extending calculation to larger systems}\label{sec:appendix3}

Using the ideas of Bloch's theorem, the numerical calculations for a finite system can be extended to a larger system (called a supercell) containing identical copies of unit cells (UC), each of size $30 \times 30$ for our case. Such a method is commonly known as `repeated zone scheme' (RZS) \cite{Ghosal02}. Here, we extend our calculation on a supercell containing $k \times k$ UC. These RZS calculations are numerically inexpensive compared to the BdG calculations on corresponding larger system and are believed to produce correct results at least for low disorder strengths when impurity-impurity correlations are weak.

We used RZS to generate a denser spectrum in the calculation of DOS and FT-LDOS by considering a supercell containing $12 \times 12$ UC.
Besides, obtaining the $q_y \rightarrow 0$ limit of $\Lambda_{xx}$ is tricky from simulations on a finite system for calculation of the superfluid stiffness. This is because of the limited $q_y$ values available on a $30 \times 30$ system from which the actual extrapolation ($q_y \rightarrow 0$) is to be made. Unlike the sSC case, where the $\Lambda_{xx}(q_y \rightarrow 0)\approx a_0+a_2 q_y^2$ ($a_0$ and $a_2$ being constants), $\Lambda_{xx}(q_y)$ shows a sub-linear behavior for a wide range of $q_y$ (not necessarily small). It is thus essential to obtain data on larger systems using RZS for an appropriate $q_y \rightarrow 0$ extrapolation. A significant numerical demand still limits $k \sim 2$ to $3$. We finally used a polynomial fit: $\Lambda^{\rm RZS}_{xx}(q_y)=\sum_p a_p q^p_y$ for $p$ up to $3$ for the final extrapolation. Fortunately, we found both in the GIMT and IMT methods that the $a_0$ is not sensitive to $p$ for the moderate to large $V$.


\bibliography{bibliography}

\providecommand{\newblock}{}

\pagebreak
\widetext
\clearpage 
~\vspace{2cm} 
\begin{center}
\end{center}
\setcounter{equation}{0}
\setcounter{figure}{0}
\setcounter{table}{0}
\setcounter{page}{1}
\makeatletter
\renewcommand{\theequation}{S\arabic{equation}}
\renewcommand{\thefigure}{S\arabic{figure}}
\renewcommand{\bibnumfmt}[1]{[S#1]}
\renewcommand{\citenumfont}[1]{S#1} 

\begin{center}
{\Large\bf Supplementary Material to ``Fate of disorder-induced inhomogeneities in strongly correlated d-wave superconductors"}
\end{center}
\vspace{0.5cm}


\vspace{0.3cm}
{ The aim of this supplementary material is two-fold: (1) Providing additional support to establish the key point made in section III C of the main paper by showing the structures in $\Delta_d(i)$ and $\rho_i$ on the lattice emphasizing their strong spatial correlations calculated using out-of-plane impurities. (2) Developing an alternative IMT scheme in place of the standard IMT described in main paper for a justified comparison with the GIMT results. The alternate IMT leads to a more meaningful comparison in certain cases. This is illustrated for FT-LDOS results, in particular.}\\[15mm]



\begin{center}
\textbf{\large Results for out-of-plane disorder}
\end{center}


An out-of-plane, screened-Coulomb potential, can be modeled as 
\begin{equation}
V_i=\sum_i V^{\rm OP}(i) \rm{exp}(-\tilde{r}_i/\lambda)/\tilde{r}_i 
\end{equation}
with $\tilde{r}_i=[(\mathbf{r}_i-\mathbf{R}_i)^2+d_z^2]^{1/2}$, where $\mathbf{R}_i+\hat{\mathbf{z}}d_z$ are the impurity locations at a vertical distance $d_z$ from the 2D plane describing the dSC, $\lambda$ is the screening length, and $V^{\rm OP}(i)$ is chosen exactly in the same manner as $V_i$ were chosen for {\it concentration disorder}. There are four independent parameters to describe the out-of-plane disorder: $d_z$, $\lambda$, $V^{\rm OP}$ and $n_{\rm imp}$. The parameters, characterizing the out of plane disorder in our calculation are: $V^{\rm OP}=1$, $n_{imp}=0.03$, $\lambda=8$ and $d_z=1.5$. 

\begin{figure}[htb]
\centering
\includegraphics[angle=0, width=0.45\textwidth]{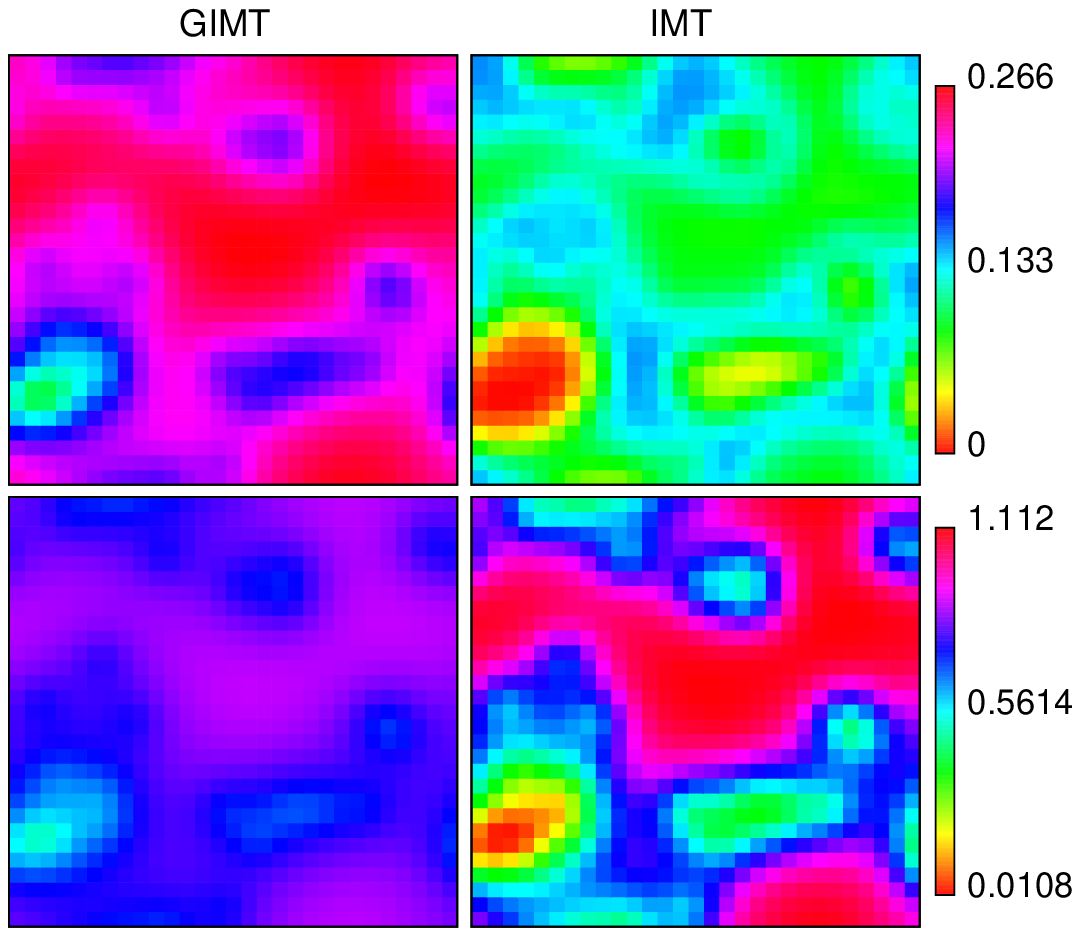}
\caption{
The spatial profile of $\rho_i$ and $\Delta_d(i)$, for out-of-plane `smooth' disorder with $n_{\rm imp}=0.03$ and $V^{\rm OP}=1$. The left column presents the GIMT results for $\Delta_d(i)$ (top) and $\rho_i$ (bottom), and the IMT results for the same appear on the right panels. While the main features are qualitatively similar to those from the box-disorder in figure~5(a) and figure~5(b) in the main paper, the extended nature of disorder in this case produces smooth variations of the order parameters. The similarity between the spatial structures of $\rho_i$ and $\Delta_d(i)$ in the GIMT results is evident! The two structures, however, are quite different in the IMT results which can be understood by looking into the regions with larger local density ($\rho_i > 1$, magenta in the bottom-right panel), which weaken $\Delta_d$ in that neighborhood. Resulting SC-`islands' are restricted only to the annular regions separating large and small $\rho_i$. In contrast, these impurities produce an impression of `islands' in the GIMT calculations, but both for $\rho_i$ and $\Delta_d(i)$.
}
\label{fig:SmoothSpat}
\end{figure}

The spatial structures of $\Delta_d(i)$ and $\rho_i$ from GIMT on the left top and left bottom of figure~\ref{fig:SmoothSpat} respectively are qualitatively indistinguishable and establish the strong spatial overlap of the large (or small) regions in both profiles. The IMT results for the same on right panel, however, show striking differences. The contrast between the GIMT and IMT results, in this regard, is sharper than the other models of disorder used in the main paper. In particular, the SC-`islands' seem to live in the annular regions supporting only a moderate local density, as expected.

\vspace{1.0cm}

\begin{center}
\textbf{\large An alternative scheme for the IMT calculations}
\end{center}

For all the comparison in the main paper, the exchange coupling $J$ was adjusted in the plain IMT, in a way to produce a homogeneous $\Delta^{(0)}_d$ same in magnitude as from the GIMT calculation with our cuprate parameters. An alternating IMT calculation possibly for a more justified comparison can be set up with the following prescription. The starting point is the minimal $t$-$J$ model without disorder, ${\cal H}=\sum_{ij \sigma}t_{ij} ({c}^{\dagger}_{i \sigma} {c}_{j \sigma}+h.c.)+\sum_{\langle ij \rangle}J_{ij}({\mathbf{S}}_i.{\mathbf{S}}_j-\frac{{n}_i {n}_j}{4})$, which only gets dressed in Gutzwiller approximation by replacing $t$ by $\tilde{t}=g^t t$, $\tilde{J}=g^{xy} J$ for the ${\mathbf{S}}_i.{\mathbf{S}}_j$ and $\tilde{J}=J$ for the ${{n}_i {n}_j}/4$ term, where the renormalization parameters $g^t$ and $g^{xy}$$=(1-\rho/2)^{-2}$ depend only on the average density, $\rho$. We get identical values for all the homogeneous order parameters, e.g., $\rho_i$, $\Delta_{ij}$ and $\tau_{ij}$ in the GIMT and IMT scheme, provided the BdG calculations are done with $\tilde{t},\tilde{J}$ in the plain IMT, while the GIMT carries the bare parameters.

Considering this clean limit as the baseline for comparison, we introduce disorder in both these calculations. We emphasize that the same strength (relative to a fixed hopping) of the disorder in the two calculations would have {\it different} absolute values: They are, $\tilde{V}=V\tilde{t}$ and $V=Vt$, where $V$ is a given strength with respect to the effective hopping for the corresponding model. We iterate that, not only the homogeneous baseline for this alternative IMT scheme are identical to those from the GIMT with respect to all the order parameters (not just $\Delta_d$ alone), it also maintains the two independent ratios of energy-scales identical, e.g., $\tilde{V}/\tilde{t} = V/t$ and $\tilde{J}/\tilde{t} = J/t$. This automatically explains the reduction of the effective disorder in the GIMT calculation, because, $\tilde{V} \sim V\times (\tilde{t}/t)$! In fact, the observation of gap filling in $N(\omega)$ for $V_0>8g^tt$ (see section III E in the main paper) motivates this alternative IMT scheme, which already incorporates the reduction of $V_{\rm eff}(i)$. Similar idea is extended to our $``t-J"$ model of equation~(2) of the main paper. 

\begin{figure}[h]
\centering
\includegraphics[angle=0, width=0.45\textwidth]{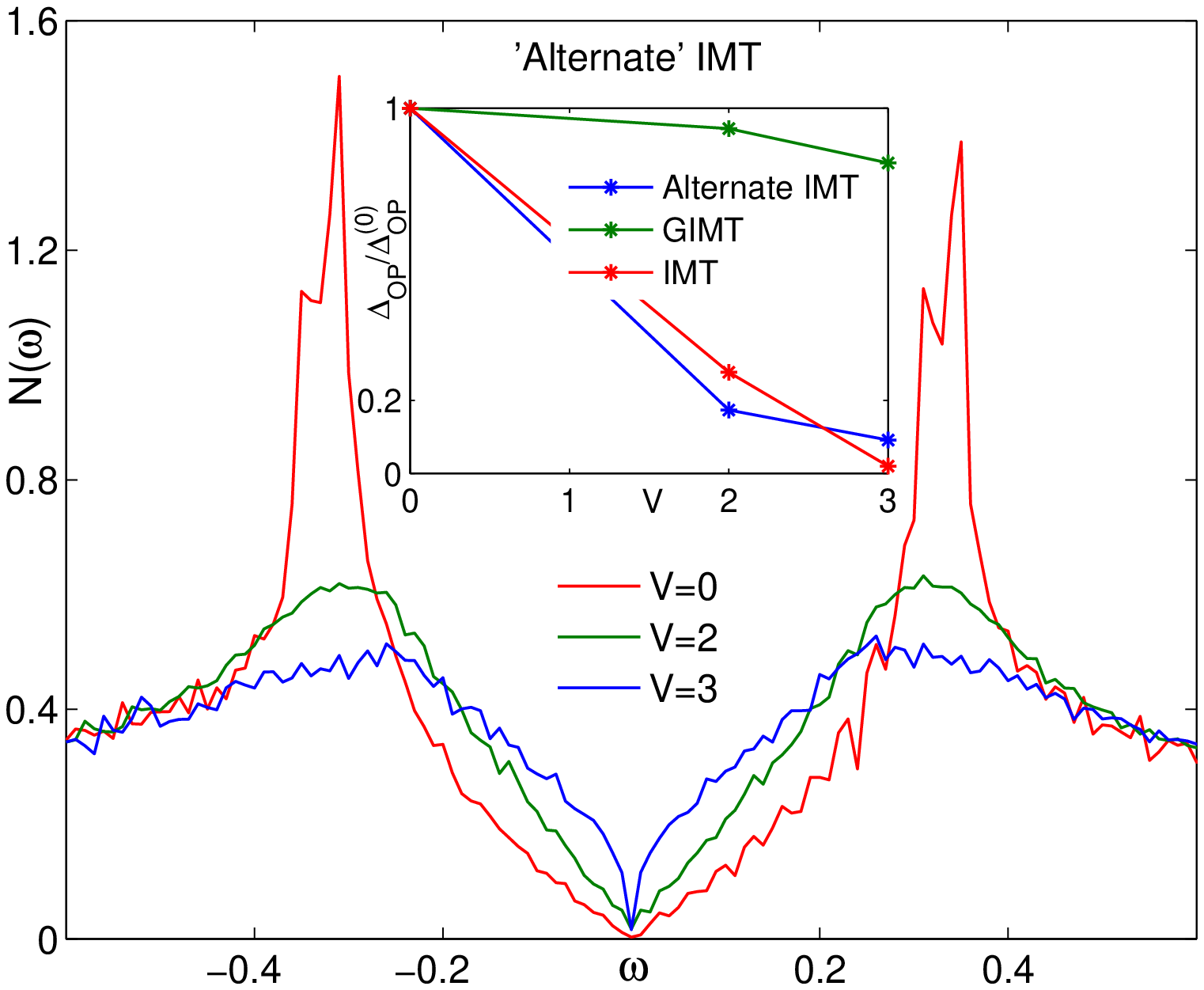}
\caption{
Site-averaged $N(\omega)$ calculated within the alternate formalism of the IMT, where the IMT-parameters are chosen such that {\it all} the order parameters be identical to those from the GIMT at $V=0$. We find the gap-filling to be weaker compared to the previous IMT findings, See figure~3(b) in the main paper. Significant differences with the GIMT results of figure~3(a) of the main paper, however, persist. The inset compares $\Delta_{\rm OP}$ from the two IMT schemes and the GIMT scheme.
}
\label{fig:altschm}
\end{figure}

\begin{figure}[h]
\centering
\includegraphics[angle=0, width=0.7\textwidth]{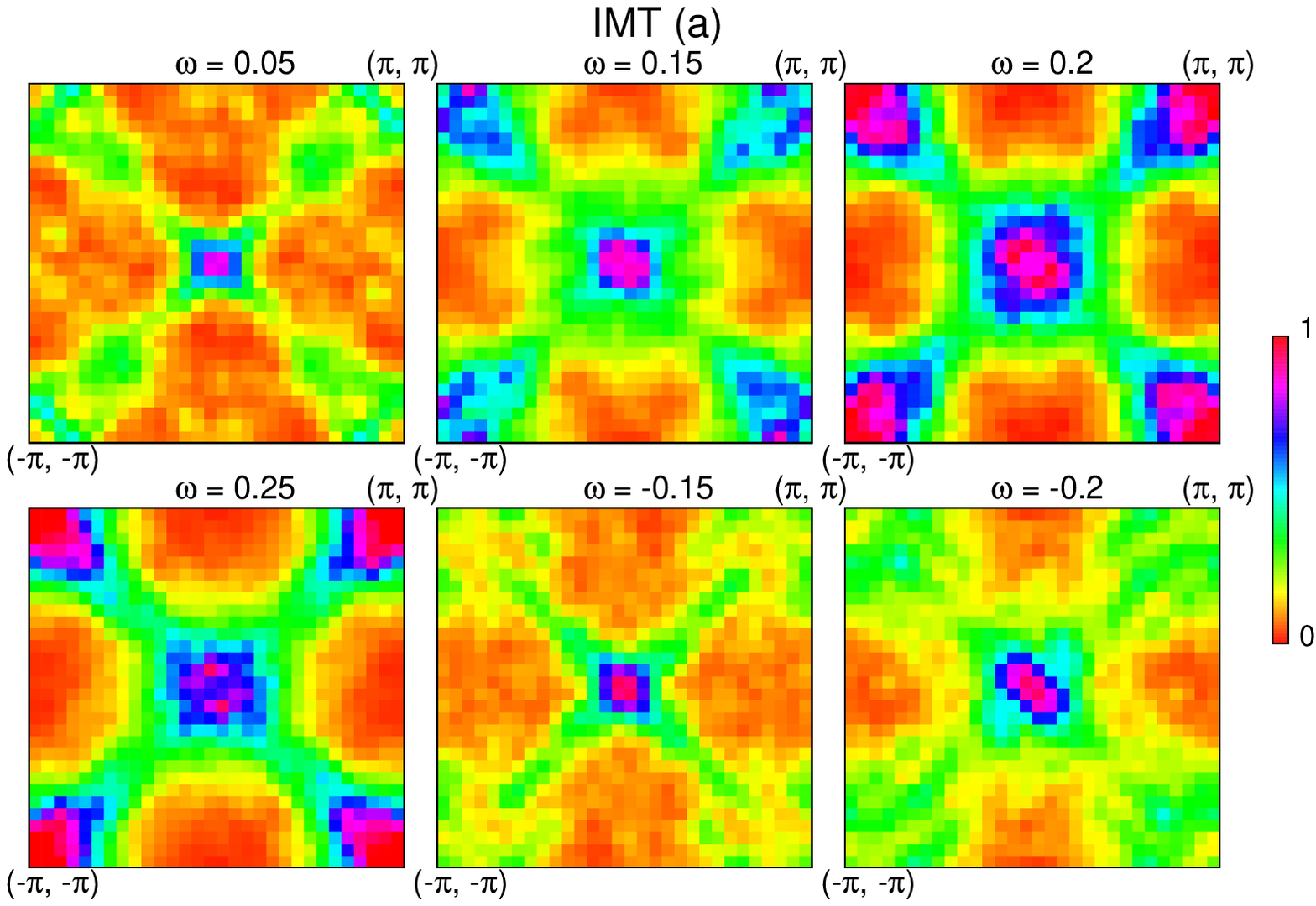}
\includegraphics[angle=0, width=0.7\textwidth]{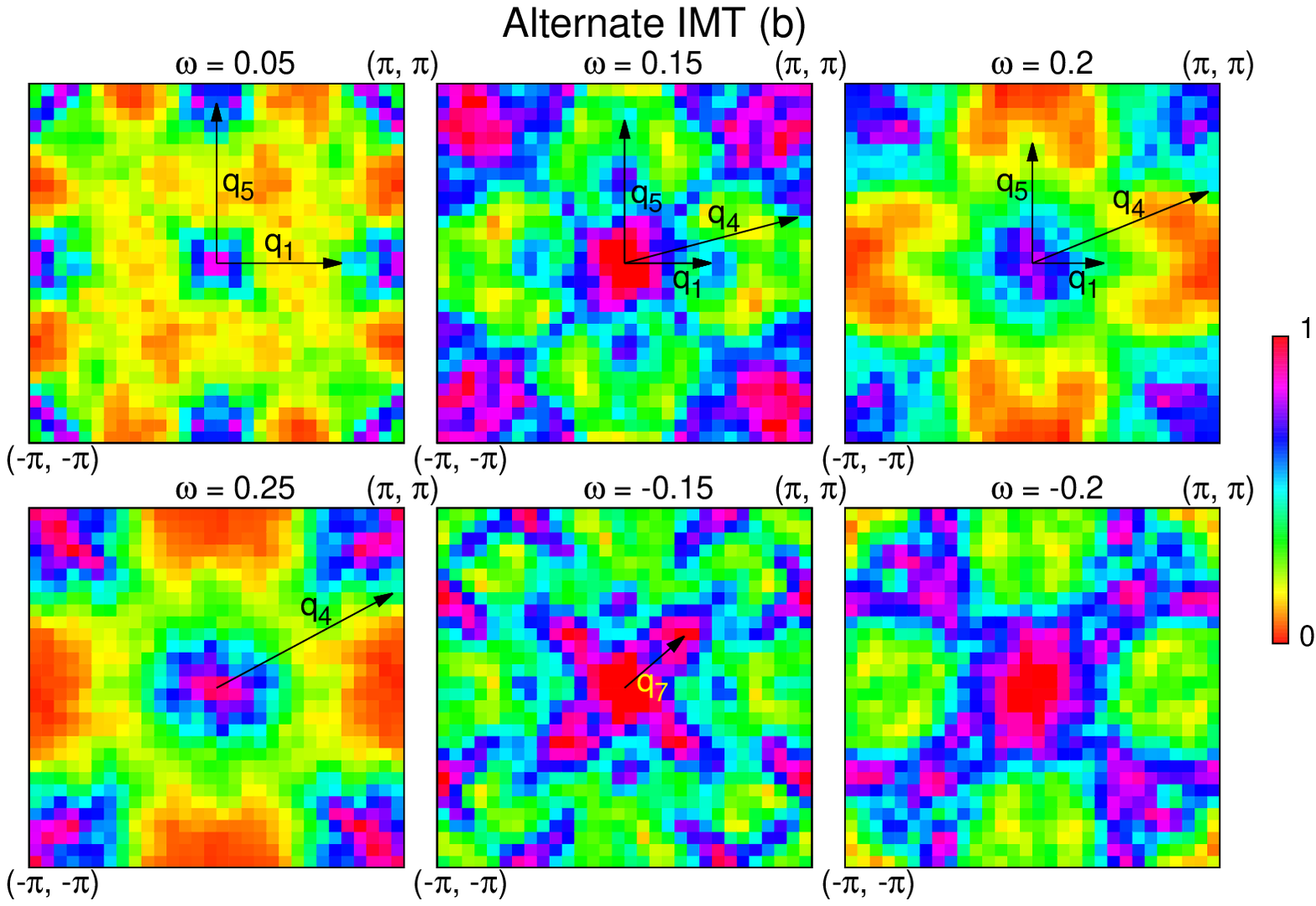}
\caption{
The color-density plot of $N(q,\omega)$ for same parameters as in figure 4 of main paper. We use the scaling such that $N(q,\omega) \in [0:1]$ for all panels. 
The maximum FT-LDOS intensites shown are truncated at different values for clarity. These are $N_{\rm max}(q,\omega)=16$ (arbitrary units) for $\omega=0.05$ and $N_{\rm max}(q,\omega)=30$ for other $\omega$ in the IMT calculations.  Alternate IMT values of similar truncations are: $N_{\rm max}(q,\omega)=7,16,48,60,20~{\rm and}~30$ for same $\omega=0.15$, $0.2$, $0.25$, $-0.15$ and $-0.2$ respectively. Almost all the octet peaks, (indicated by arrows) which are visible in figure 4 (main paper), are also visible in (b). 
}
\label{fig:ftldos}
\end{figure}

We repeated most of our calculations in the main paper with this alternative formalism. The differences between the GIMT and IMT results reduce somewhat with this alternative IMT calculations. The reduction is significant for $N(\omega)$ and FT-LDOS, so that these deserve a detailed discussion. The results for $N(\omega)$ are summarized in the figure~\ref{fig:altschm} (compare with figures~3(a) and 3(b) of the main paper). Differences in $\Delta_{\rm OP}$ still remain large, as can be seen from the inset of figure~2. The filling-up of low-lying $N(\omega)$ is found weaker compared to the standard IMT scheme (of main paper) in figure~3(b) (main paper), yet the averaged DOS is not as robust as it is from the GIMT results in figure~3(a) (main paper).

FT-LDOS too reflects significant differences between the standard IMT scheme (in the main paper) and the alternative IMT scheme (of current discussion), as presented in figure~\ref{fig:ftldos}. While none of the FT-LDOS peaks stands out in the standard IMT results (see figure~\ref{fig:ftldos}(a)) making them significantly different from the GIMT findings (see figure~4 in the main paper), the same results from the alternate IMT recover much of the signatures of the GIMT findings. The peaks of figure~\ref{fig:ftldos}(b) and those in figure~4 (main paper) appear at the same locations, though small quantitative mismatches persist. For example, the high momentum features at the Brillouin Zone corners in figure~\ref{fig:ftldos}(b) is stronger than the GIMT results. In contrast, none of the FT-LDOS peaks stands out in the standard IMT results (figure~\ref{fig:ftldos}(a)). We believe that the differences in the two IMT schemes stem from the $\omega$ values, because, in the alternate IMT scheme $\omega$ is measured in terms of $g^t t$ while for IMT the scale is just the bare $t$.
The similarity of the relative features of figure~\ref{fig:ftldos}(a) and of figure~4 (main paper) hints that the periodic and energy resolved modulation of the local density in cuprates might not originate just from the strong correlations.

We conclude that, while the renormalization of effective disorder (taken into account in the alternating IMT scheme, by construction) plays a significant role to differentiate the GIMT and IMT results, there are more to it, likely related to the complex interplay of different order-parameters through a complete self-consistency.

\end{document}